\documentclass[lettersize,journal]{IEEEtran}
\IEEEoverridecommandlockouts

\usepackage{amsmath,amsfonts,amssymb,amsthm,amstext}
\usepackage{bbm}

\usepackage{graphicx}
\usepackage{textcomp}
\usepackage{xcolor}

\usepackage{textcomp}
\usepackage{wrapfig}

\usepackage{algorithmic}
\usepackage[ruled,vlined,linesnumbered]{algorithm2e}

\SetCommentSty{mycommfont}

\usepackage{array}

\usepackage{booktabs}
\usepackage{caption}
\usepackage{cite}
\usepackage{subcaption}
\usepackage{enumerate}
\usepackage{float}
\usepackage{epstopdf}
\usepackage{stmaryrd}
\usepackage{multirow,epsfig}
\usepackage{mdwmath}
\usepackage{mdwtab}
\usepackage{marvosym}
\usepackage{hyperref}
\newtheorem{lem}{Lemma}
\newtheorem{mydef}{Definition}
\newtheorem{mytheo}{Theorem}
\newtheorem{myexam}{Example}

\begin{document}

\title{\pmb{\texttt{RASE}}: Efficient Privacy-preserving Data Aggregation against Disclosure Attacks
for IoTs}

\author{Zuyan Wang{\IEEEauthorrefmark{2}\IEEEauthorrefmark{3}\IEEEauthorrefmark{4},
Jun Tao \IEEEauthorrefmark{2}\IEEEauthorrefmark{3}\IEEEauthorrefmark{4}, 
Dika Zou\IEEEauthorrefmark{2},}\\
\IEEEauthorblockA{\IEEEauthorrefmark{2}School of Cyber Science and Engineering, Southeast University, Nanjing, China\\
\IEEEauthorrefmark{3}Key Lab of CNII, MOE, Southeast University, Nanjing, China\\
\IEEEauthorrefmark{4}Purple Mountain Laboratories for Network and Communication Security, Nanjing, China}\\
\{zywang92, juntao, dikaizou\}@seu.edu.cn
}

\markboth{Journal of \LaTeX\ Class Files,~Vol.~14, No.~8, August~2021}%
{Shell \MakeLowercase{\textit{et al.}}: A Sample Article Using IEEEtran.cls for IEEE Journals}
\maketitle

\begin{abstract}
The growing popular awareness of personal privacy
raises the following quandary:
what is the new paradigm for collecting and protecting the data produced by ever-increasing sensor devices.
Most previous studies on co-design of
data aggregation and privacy preservation
assume that a trusted
fusion center adheres to privacy regimes.
Very recent work has taken steps towards relaxing the
assumption by allowing data contributors 
to locally perturb their own data.
Although
these solutions
withhold some data content to mitigate privacy risks,
they have been shown to offer insufficient
protection against disclosure attacks.
Aiming at providing a more rigorous data safeguard
for the Internet of Things (IoTs), this paper initiates the study of
privacy-preserving data aggregation. 
We propose a novel paradigm (called {\texttt{RASE}}),
which can be generalized into a 3-step sequential procedure\,--\,noise addition, followed by random permutation, and then parameter estimation.
Specially,
we design
a differentially private randomizer,
which carefully guides data contributors to obfuscate the truth.
Then,
a shuffler
is employed  to
receive the noisy data from all data contributors.
After that,
it breaks
the correct linkage between senders and receivers by applying a random permutation.
The estimation phase involves using inaccurate data to calculate an approximate aggregate value.
Extensive simulations are provided to
explore the privacy-utility landscape of our {\texttt{RASE}}.
\end{abstract}


\section{Introduction}
\label{sec:Introduction}
\begin{figure}
  \centering
  \includegraphics[width=.45\textwidth]{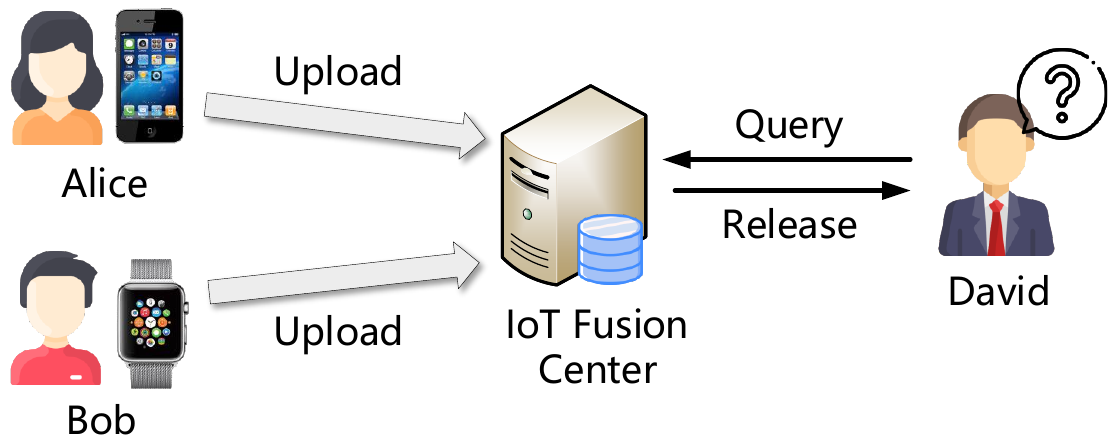}
  \caption{An example of the system model.}\label{fig:Basic Design}
\end{figure}

The Internet of Things, or IoTs,
is an intelligent ecosystem
of various web-enabled devices~\cite{chettri2020comprehensive}.
These IoT devices
collect massive amounts of granular data about individuals'
habits and activities,
and thus become attractive targets for malicious attacks.
Most importantly,
with the evolution of privacy and data protection laws
such as the progressive EU General Data Protection Regulation~\cite{regard2013recommendation},
there has been a rising societal concern around how and when personal data are used~\cite{lupton2016personal}.
This paper considers the following scenario often seen in our daily life.
Suppose that Alice and Bob wish to teleport the exercise data to third-party applications for deeper understanding of their physical well-being. Apart from health needs,
they hold a paramount concern for their personal privacy.
Meanwhile, David, an employee at a company subscribing to the application platforms, is tasked with analyzing statistics to identify areas for product innovation. 
A visualization of the system model is shown in Fig.~\ref{fig:Basic Design}.
We can not guarantee that the data or the statistic is privately accessible only to Alice and Bob.
There is always the risk that an entity acquires the sensitive information through other means\,--\,from the platform recording all personal history, to David engaging in hacking activities.
The mentioned privacy issue is also quite common in many
IoT applications,
e.g.,
smart homes~\cite{rizi2022systematic},
precision agriculture~\cite{shu2021guest} and so on~\cite{sun2019relationship, wang2021privacy, xiong2019personalized}.
Therefore,
ensuring data privacy has emerged as an urgent task
for IoT-related services.
We should seek a paradigm to safeguard the personal data from any form of exploitation or disclosure. 

\emph{Differential privacy}~\cite{Qian2017Privacy, Dwork2016Calibrating, Dwork2013The,bao2015ddpft}
is the celebrated privacy concept
that comes with provable privacy measure,
independent of the computing ability available to an adversary.
As a result, it is frequently employed in IoT data aggregation to guard privacy.
Early studies
concentrated on
designing and implementing differential privacy-based data aggregation within the centralized model,
where privacy is only guaranteed at the end of data analysis process.
Subsequent work has expanded the definition to a distributed model.
Different from the setting of the centralized model,
in the distributed model,
the data contributor trusts no one else
and obtains strong privacy during the data collection process.
Either way,
the general paradigm of using differential privacy
in data aggregation
is
to suitably perturb the original data to ensure that the releasing results are statistically indistinguishable.

\subsection{Problem Description and Motivation}
Unfortunately, the perturbation-driven paradigm does not make strong assumptions about the background knowledge of adversaries.
In consequence, it still faces vulnerability to disclosure attacks~\cite{liu2018generalized},
especially \emph{identity disclosure}.
Advanced adversaries may leverage external knowledge to de-anonymize the sender's identity. 
For the above example,
ill-intentioned third parties
might be able to link the received data to
the device's IP address,
enabling them to track more traffic information about Alice and Bob,
such as routing details, network activities, and communication patterns.
This knowledge, in turn, could be used to reverse-engineer the process of noise addition, ultimately unveiling the perturbation mechanism and recovering the unaltered data.
Such a breach poses a grave threat to personal privacy. 

Moreover,
the
efficacy of the aggregated data for
practical applications hinges significantly on data accuracy,
which means ``blind'' noise addition may negatively degrade the quality of services.
On the other hand, 
without a sufficient amount of noise, the data could be decrypted with a high probability. 
To illustrate, consider the context of location-based services.
If the noise injected to the true location is an overkill,
the utility of the data will severely diminish.
This could cause higher-than-expected errors when making decisions about resource allocation and providing location-based recommendations.
Such inaccuracies could have far-reaching consequences, even extending to system stability. 
Consequently, there is an ongoing imperative to balance privacy protection and data usability.

The above observations motivate us to pose
the problem\,--\,\emph{can we pursue customized
data utility while suppressing
disclosure attacks?}
Ideally, we can achieve privacy amplification along with low privacy cost. 
However, due to inherent limitations
(e.g., computational resources, storage capacity and battery power), directly deploying such complicated  technologies on resource-limited IoT devices is a challenging choice.
To tackle the problem,
we must seek an effective solution that
is based on the hybrid model of privacy protection.

\subsection{Our Contributions}
In this paper,
a \emph{randomize-shuffle-estimate} paradigm (we name it ``\pmb{\texttt{RASE}}'')
is proposed to aggregate the sensory data from privately-held IoT devices.
The structure of \pmb{\texttt{RASE}} falls between the distributed model and the centralized model.
There are
a number of aggregation operations
widely used in statistical data analysis.
For the sake of clearness,
we narrow our focus on
the average aggregation,
but our \pmb{\texttt{RASE}} is not restricted to this; it can be applied to any other aggregation methods as well.
Here is a breakdown of \pmb{\texttt{RASE}}:
(\romannumeral 1)
the fusion center
decides how much accuracy it needs and  broadcasts the requirements to the data contributors;
(\romannumeral 2) the randomizer is a local  function that combines these requirements to calibrate noise;
(\romannumeral 3)
the shuffler applies a given permutation to hinder identity identification; and (\romannumeral 4)
the estimator is utilized to calculate an approximation of the mean average.
By implementing \pmb{\texttt{RASE}},
the data privacy is preserved against
the disclosure attacks,
and at the same time the data accuracy is under a certain bound.

The novelty of this paper lies in that we highlight the privacy-utility challenge associated with data aggregation in IoT-related applications. We develop \pmb{\texttt{RASE}}, a novel mechanism that achieves better trade-offs via analyzing specific requirements. 
\pmb{\texttt{RASE}} complies with all the interested dimensions and offers salient features, including low complexity and provable privacy guarantees. To the best of our knowledge, we are the first in-depth work to advocate the use of random perturbation for preserving the data collection process and the integration of randomly generated permutation into obfuscate identities.

Our main contributions are three-fold:

\begin{itemize}
  \item We develop dynamic feedback from the fusion center to standardize the data error range. We design a budget-aware local randomizer~\texttt{BR} and a robust shuffler~\texttt{RS}, both of which serve as the basic building block of \pmb{\texttt{RASE}} to achieve privacy protection. The \texttt{BR} randomizer is a variant of the Laplace mechanism. It incorporates a noise clamping technology on each output of Laplace noise to filter out excessive randomness. Our shuffler~\texttt{RS} divides the data sequence into fragments in disorder and  then re-links them, thereby improving sender anonymity. We showcase that \texttt{RS} works even for the extreme cases. 
  \item We introduce several estimators to approximate the mean average of the aggregated data. First, the sample mean estimator is the common way to obtain an unbiased estimate, but it is susceptible to the presence of outliers.
  Second, the maximum-likelihood estimator minimizes the likelihood function over   the contaminated data and often generates the most accurate estimates. Third, the bootstrap estimator is an iterative procedure, which generates estimates through data resampling. 
  We exemplify the use of the mentioned estimators in a five-element set and provide a qualitative comparison with them.  
  \item We conduct a comprehensive experiment on a real data stream collected from the REFIT smart home dataset~\cite{Firth2017REFIT}. The experimental results demonstrate the superiority of our \pmb{\texttt{RASE}}.
  For example,
  our new paradigm can make a more appropriate trade-off between utility and privacy than the state-of-the-art algorithms.
\end{itemize}

The remainder of this paper is organized as follows.
Section~\ref{sec:Preliminaries} introduces the technical backgrounds and Section~\ref{sec:System} formulates the data aggregation problem in IoTs.
Our sanitization framework
is described in Section~\ref{sec:Solutions}.
The computational results are given in Section~\ref{sec:Evaluation}.
The related literature is discussed in
Section~\ref{sec:Related Work}.
We
conclude the paper in Section~\ref{sec:Conclusion}.
\section{Preliminaries}\label{sec:Preliminaries}
In this section, 
we delineate some necessary notations and definitions for our next work.

\subsection{Local Differential Privacy}
The concept of local different privacy (LDP)~\cite{Kasiviswanathan2008What, Graham2018Privacy}
was recently presented for measuring the privacy guarantees that a distributed algorithm has provided.
Unlike the traditional differential privacy~\cite{Dwork2016Calibrating, Dwork2013The},
where users are uncomfortable sending their  raw data to an aggregator directly,
LDP comes into play in the data collection process.
Specifically,
under the local setting of differential privacy,
each user locally perturbs their data via
injecting random noise
before sharing it.
By taking full control of data purifying in their own hands,
individuals can keep sensitive information private even if the aggregator is malicious.
Benefiting from this advantage,
LDP has been widely implemented in many real-world applications,
especially
in the situations where the trust relationship is difficult to establish.
In the following, 
we
provide a formal introduction for LDP.
\begin{mydef}[$\epsilon$-local differential privacy~\cite{Graham2018Privacy}]\label{def:LDP}
A randomized mechanism $\mathcal{M}$ satisfies $\epsilon$-local differential privacy (or $\epsilon$-LDP),
if for any pair of values $x$, $x'$,
and $Y \subseteq {\rm range}(\mathcal{M})$,
\begin{equation}
 {\rm Pr}\Big[\mathcal{M}(x)= Y\Big]\leq e^{\epsilon}\cdot {\rm Pr}\Big[\mathcal{M}(x')= Y\Big],
\end{equation}
where the probability is
taken
over the randomness of $\mathcal{M}$.
\end{mydef}

In Definition~\ref{def:LDP},
the system parameter $\epsilon$ is often referred to as the privacy budget,
by which a single individual can adaptively adjust the privacy-preserving strength of mechanism $\mathcal{M}$.
Conceptually,
we give this explanation:
a smaller value of $\epsilon$ results in stronger privacy preservation,
while a higher value of $\epsilon$ means more similarity between neighboring data but less privacy guarantees.
LDP also has the post-processing property:
if the mechanism $\mathcal{M}$ satisfies $\epsilon$-LDP,
then for every deterministic function $g(\cdot)$,
$g(\mathcal{M})$ is at least $\epsilon$-LDP.
This means that 
employing some of these post-processing steps will not reverse the privacy guarantees provided by the differentially private mechanism\,---\,there is a possible tendency to be more time consuming than no  post-processing.  

\begin{figure}
  \centering
  \includegraphics[width=.43\textwidth]{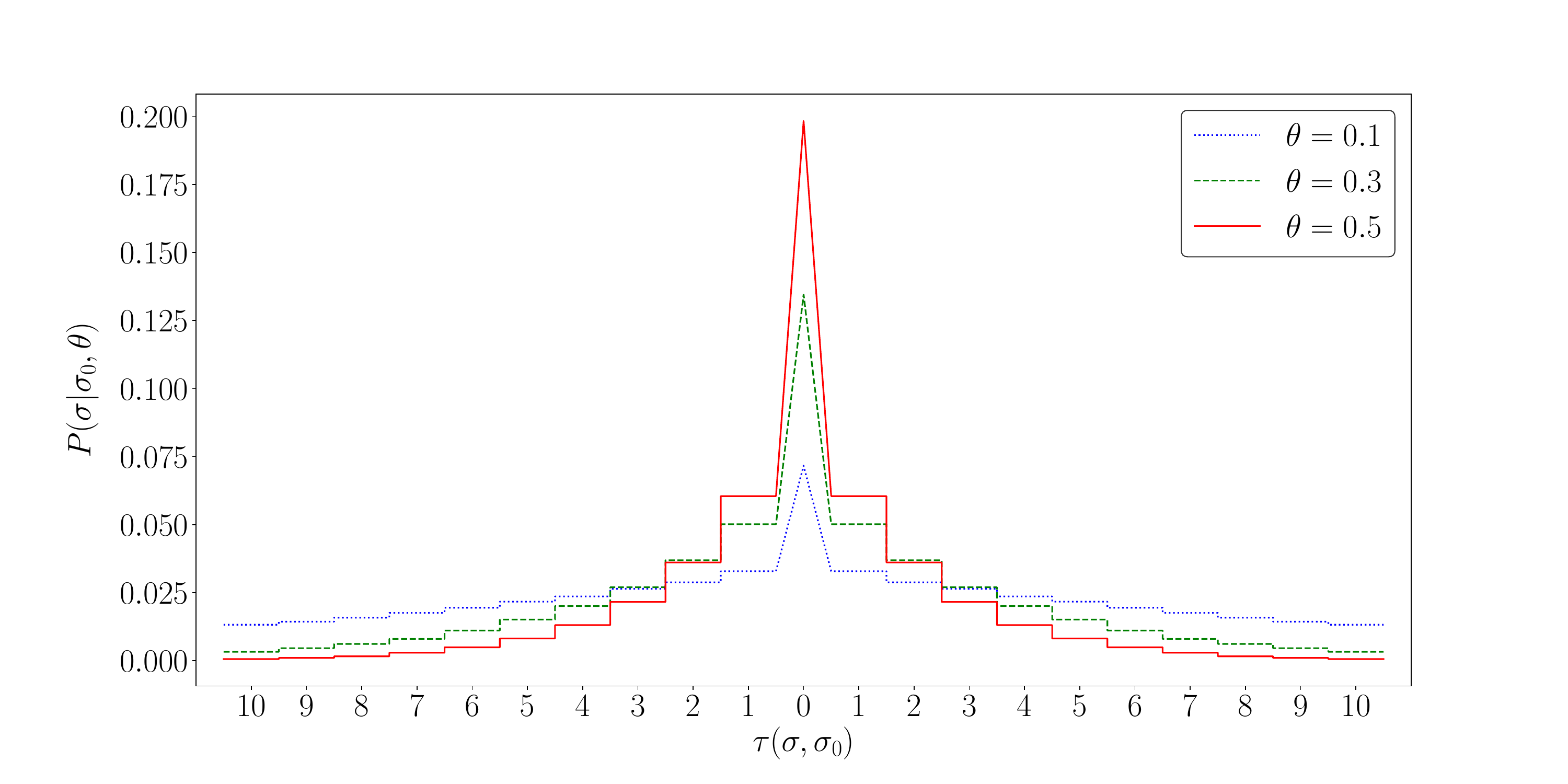}
  \caption{When $m=5$,
the probability of the Mallows model at each Kendall tau distance for different $\theta$.}\label{fig:Mallows}
\end{figure}

\subsection{Mallows Mechanism}
In mathematics,
a permutation (or ordering) is
a bijection of the set of
integers $\llbracket n \rrbracket=\{1,\ldots, n\}$ onto itself.
The set of all $n!$ possible
permutations forms a
symmetric group,
denoted by $S_n$.
For example,
there are six permutations of the three-element set $\{1, 2, 3\}$,
namely,
$(1, 2, 3)$, $(1, 3, 2)$, $(3, 1, 2)$, $(3, 2, 1)$, $(2, 3, 1)$, and $(2, 1, 3)$.
Commonly,
we will use lowercase Greek letters $\sigma$, $\tau$ as permutations,
boldface type $\boldsymbol{\sigma}(i)$ as the value of index $i$,
and $\boldsymbol{\sigma}^{-1}(i)$ as the assigned index of item $i$.
The permutation that places each term in numerical order is called the identity permutation and denoted as $\sigma_I=\big(1, \ 2,\ \ldots, \ n\big)$.
We will need the next definitions of permutations for our goal.

\begin{mydef}[Compositionality, associativity and invertibility]\label{def:pro_per}
Given any two permutations $\sigma$ and $\sigma'$ in symmetric group $S_n$,
the following holds:
\begin{itemize}
  \item Their composition $\sigma \sigma'$ is also in $S_n$  (compositionality).
  \item   $(\sigma  \sigma')\tau = \sigma  (\sigma' \tau)$ for any $\tau \in S_n$ (associativity).
  \item  For every $\sigma$,
  there exists an inverse permutation $\sigma^{-1}\in S_n$ such that $\sigma^{-1}\sigma=\sigma_{I}$ (invertibility).
\end{itemize}
\end{mydef}

\begin{mydef}[$d_{\sigma}$-privacy~\cite{Meehan2021Privacy}]\label{def:d_sigma}
   A randomized mechanism $\mathcal{A}$ satisfies $d_{\sigma}$-privacy if for any pair of permutations $\sigma$, $\sigma'$ and  $\tau\subseteq {\rm range}(\mathcal{A})$,
 \begin{equation}
    {\rm Pr}\Big[\mathcal{A}(\sigma)= \tau\Big]\leq e^{\alpha}\cdot {\rm Pr}\Big[\mathcal{A}(\sigma')= \tau\Big],
 \end{equation}
where the probability is
taken over the randomness of $\mathcal{A}$.
\end{mydef}

The Mallows model~\cite{Mallows1957Non} is a
parametrized,
distance-based probabilistic model over
permutation spaces.
Formally,
it can be formulated as
\begin{equation}\label{eq:mallows}
{\rm Pr}\Big[\sigma\mid \sigma_0, \theta\Big]=\frac{{\rm exp}\big({-\theta \cdot d(\sigma,\sigma_0)}\big)}{Z(\theta)},
\end{equation}
where $\theta$ and $\sigma_0$ are the parameters of the model:
$\sigma_0\in S_n$ is the central permutation,
$\theta$ is the spread parameter,
and $Z(\theta)$ is the normalizing constant.
Additionally,
$d(\cdot,\cdot)$ is the distance measure metric,
which for our paper will be the Kendall tau distance.
That is,
the number of pairaware order disagreements
$d_K(\sigma,\sigma_0)=\sum_{1\leq i\leq j \leq m}\mathbbm{1}\{ (\boldsymbol{\sigma}(i)-\boldsymbol{\sigma}(j))(\boldsymbol{\sigma}_0(i)-\boldsymbol{\sigma}_0(j))\}$,
where $\mathbbm{1}(\cdot)$ denotes the indicator function.
The corresponding distribution is illustrated in Fig.~\ref{fig:Mallows}.
It is obvious to see that,
when $\theta \geq 0$,
(\romannumeral 1)
$\sigma_0$ is the mode of the distribution with the highest probability,
and (\romannumeral 2)
the probability of the other $(5!-1)$ permutations decays exponentially with the distance to $\sigma_0$.
These two properties show that Mallows model is somewhat analogous to the Gaussian distribution,
thereby belonging to the family of exponential
distributions.

This model has
been used in the literature to randomly shuffle the permutations,
which normally refers to as the Mallows mechanism~\cite{Meehan2021Privacy}.
The key idea of the mechanism
is to 
sample a permutation from the Mallows distribution,
and then to apply the random permutation to
rearrange the ordering of original one.
The design of our identity hiding mechanism stems from
this the random permutation
generator.

%
\section{Problem Formulation}
\label{sec:System}
In this section,
we give the details of our privacy problem by formalizing the system model and the threat model.
\subsection{System Overview}\label{subsec:system}

As shown in Fig.~\ref{fig:Basic Design},
the system consists of the
following three entities:
(\romannumeral 1) data contributors,
(\romannumeral 2) a fusion center
and (\romannumeral 3) subscribers.
Assume that there are $n\geq 2$ data contributors $u_i$, $i\in \llbracket n \rrbracket$, 
each with a personalized privacy specification $\epsilon_i$.
We describe the correlations
among these data contributors by using an undirected connected graph $G=(V,E)$,
where $V$ denotes the vertex set containing $n$ data contributors and $E \subseteq V\times V$ denotes the edge set.
An edge $(i, j)\in E$ implies that data contributors $u_i$ and $u_j$ can interact with
each other.
Then,
we have a dedicated number of groups
denoted by $\{G_1,\ldots, G_{k}\}$.
Each group is disjoint and includes those data contributors who are ``closest'' in the graph connectivity.

In our setting, 
data is assumed to collect by the   
fusion center with the consent of the $n$ data contributors
at regular intervals.
For each data source $u_i$,
we define the data format as a triplet in the
form of $x_i^{(t)}=(State, ID, Time)$,
namely that it is associated with the state observed from an unique device
at a given point in time.
For the sake of clearness, 
we represent time with logical timestamps by 
letting $t = 1, 2, 3, \ldots, T$. 
Taking location-based systems as an example,
$x_i^{(t)}$ could be the
geographic information on $u_i$'s mobile device over time intervals of one second.
The subscribers act as data consumers and request the data for their
own purpose. 

\noindent{\bf Aggregate Queries on Data.}
In fact,
a plethora of aggregation operations 
can be used to conduct data integrating,
but here we focus on the average aggregation,
which reflects the average of observations happening
at the same timestamp. 
More formally,
the average-case analysis in the collected data is
defined as
\begin{equation}\label{eq:aver_val}
    \overline{x}^{(t_q)}=\frac{1}{n}{\sum}_{i\in\llbracket     1,n \rrbracket} x_i^{(t_q)},
\end{equation}
where $t_q$ denotes the timestamp of when  subscribers send a request to the fusion center. 
We can also enrich the basic query through supporting time-based
windows.
Let $w$ denote 
a window size.
The query ${\rm Win}(t_q,w)$ is the average value of all states within
time interval $\llparenthesis t_q-w,t_q\rrbracket$.
From Eq.~(\ref{eq:aver_val}), 
we can derive ${\rm Win}( t_q,w)=\frac{1}{w}\sum_{t=t_q-w+1}^{t_q}\overline{x}^{(t)}$.
To clarify the targeted system,
we illustrate with an example.

\begin{myexam}\label{exa:sys}
Referring back to Fig.~\ref{fig:Basic Design},
consider that Alice and Bob wear IoT-enabled devices
to estimate the total calories burned and track the daily physical activity.
All the performance data are decoded by the fusion center 
to extract insights on their health condition.
David pays for access to the system 
and wants to get the statistics that might be used to promote products.
Obviously, the data are at risk of being sold for making money.
Due to safety concerns,
we need to ensure 
that only the sanitized data can flow in the scenario.
\end{myexam}

\subsection{Threat Model and Security Assumption}\label{subsec:threat}
In general, 
the goal of an adversary is to 
infer the
proprietary information of individuals.
From the above system settings,
we can see that the fusion center and the subscribers
have the upper hand,
because they always have a chance (albeit possibly
small) of getting the input data.
Consequently,
we investigate the
privacy issue
at either
the fusion center or subscribers.

\noindent \textbf{Fusion Center's Attacks.}
Similar to Tran ~\cite{Tran2022Smart},
we consider the assumption that,
roughly speaking,
the IoT fusion center 
has fragmentary knowledge about the system entities and behaves semi-honestly during whole aggregation phase.
That is,
the basic functional units of the fusion center are persistently made available to each party,
but privacy of data contributors may be violated nevertheless.
In this vein,
the fusion center can
identify
the sender of data
(regardless
of data contributor's privacy settings),
and can also
monitor who is accessing which fraction of the stored sensory data.

\noindent \textbf{Subscriber's Attacks.}
Like \cite{Qian2017Privacy}, \cite{Chen2012Towards} and \cite{Huang2018Reliable},
we assume
that subscribers do not collude with
other entities to obtain private information about the data contributor.
The reason is that
if doing this they
have a high risk of being reported and exposed by other participants.
Moreover,
the subscribers
can not interfere with the system by
adding,
removing or modifying the data on the fusion center.
Their malicious operations would include (\romannumeral 1) learning data contributor's sensitive attributes (e.g, home addresses and medical records) from generally-sensitive data,
(\romannumeral 2) creating behavioral profiles that can be sold,
or (\romannumeral 3) eavesdropping on the data flow about points of interest.


\subsection{Problem Definition}
In this paper,
we study the following privacy issue:
\emph{given a group of data contributors, 
each of who has private data and a privacy specification,
an adversary aims to 
reveal sensitive information,
like identities and other properties,
our goal is to design an efficient
mechanism to suppress realistic
privacy disclosure while retaining meaningful data utility.}

%
We desire that our solution has the technical perspectives 
from the view of
both data privacy and utility.
\underline {\textit {Data
Privacy:}}
this 
concentrates on
the protection of data from misuse,
including means how effectively
a privacy mechanism can
confuse the adversary.
\underline {\textit {Data Utility:}}
typical
user-centered information services
are based on the learnability of input data;
it is of paramount importance to
fulfil the utility
requirements of service providers. 
These two perspectives go in opposite directions\,---\,the higher the privacy protection level is required,
the lower the accuracy will be.
To make a trade-off,
we theoretically and experimentally
explore the privacy-utility landscape
in the remaining sections.

\section{Our Countermeasure}
\label{sec:Solutions}
\begin{figure}
  \centering
   \includegraphics[width=.48\textwidth]{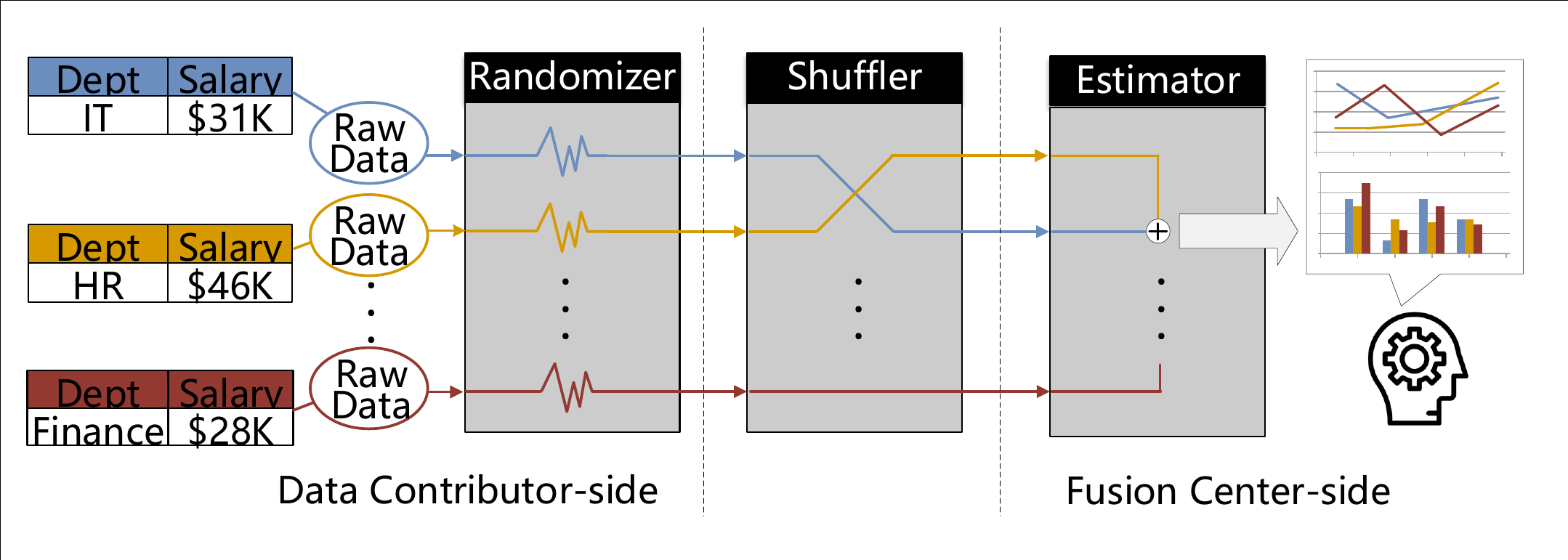}
  \caption{Overview of \pmb{\texttt{RASE}}: sanitization of sensitive data with a three-part privacy preservation framework.}\label{fig:overview}
\end{figure}
In this section,
we present a comprehensive paradigm (called \pmb{\texttt{RASE}}) for
aggregating data from multiple sources.
Fig.~\ref{fig:overview} illustrates 
the key components of our method:
(\romannumeral 1) the randomizer  takes input from a data contributor, injects noise to each
sensed value, and releases a message of the noisy result;
(\romannumeral 2) the shuffler is abstracted as a mediator that obscures the identifying information from the messages before they are handled to the fusion center; 
(\romannumeral 3) the estimator 
deals with truth discovery from the received messages.
The combination of randomizer and shuffler forms our line of defense,
while the estimator provides insight on how to optimize data utility under randomness.
For the sake of clarity,
we start by sketching the role of each building block in \pmb{\texttt{RASE}}.
Finally, 
we show how to integrate them into one unified.



\subsection{Design of Randomizer}\label{subsec:ran}
We first design a local randomizer, known as \texttt{BR}, which allows the data contributor to achieve the desired level of security
while being aware of the precision requirement of the fusion center.
To enforce LDP,
the perturbation-based idea
is widely adopted in the literature, e.g.,~\cite{Wang2021Arbitrarily, Qu2021An, Tran2022Smart},
in which
an appropriate amount of random noise is applied to perturb the data.
We also adopt this idea 
to generate noise.
We note that the data $x_i^{(t)}$ varies at every timestamp,
and therefore 
may be unbound.
We assume that the tuples of $x_i^{(t)}$ are independent with each other.
When clear from context,
we will drop superscript $(t)$ and just use $x_i$ for the sensed value.


 %
%
%
%
%
%
%

The adjacent
relation is one of the important concepts
in perturbation.
It describes how much
``variability'' a set of data is.
Here, the variability,
also called dispersion,
is measured by the range of data.
In practice,
the sensory data of the same IoT applications could remain constant within a specific range, 
due to some physical characteristics.
For example,
the measurement result of temperature sensors always ranges from $-40$ to $125$ degrees Celsius.
Thus,
we assume that the data of each 
$u_i$ take values in the same range $[x_{min}, x_{max}]$
and let $\Delta(x)=x_{max}-x_{min}$ for $\forall i\in \{1,2, \ldots, n \}$.
Here, $x_{max}$ is largest value of the sensed data and $x_{min}$
is the smallest.
In light of the fact that 
the learnability of data is required by the fusion center,
the added noise should be maintained at a certain level.
Towards this end,
we propose a definition of the interval precision
to quantify how accurate the obfuscation is in preserving privacy.

\begin{mydef}[Interval precision]\label{def:accuracy}
Let $\beta$, $\rho\in[0,1]$. The noisy version $y_i$ of data $x_i$ satisfies  $(\beta, \rho)$-determined interval precision if the following inequality holds:
\begin{equation}\label{ineq:accuracy}
{\rm Pr}\Big[y_i\geq (1-\beta)x_i~{\rm and}~ y_i\leq(1+\beta)x_i  \Big]\geq \rho \ . 
\end{equation}
\end{mydef}
Intuitively,
this definition
indicates that 
the noisy result $y_i$
lies within a range of the original data $x_i$,
with a probability $\rho$.
From the perspective of estimation theory,
$\beta$ describes the confidence interval and $\rho100\%$ represents
the confidence level.
For a given $\rho$,
a smaller $\beta$ means a better data accuracy.
With this fact,
it is possible to leverage the confidence interval under a certain
confidence level for limiting the influence of excessive noise.
Assume that all data contributors take the same privacy protection strength $\epsilon_s$. 
We fine tune the Laplace mechanism~\cite{Dwork2016Calibrating}
as the
implementation of the randomizer.
The following lemma can
guide us to regulate the Laplace noise.
\begin{lem}\label{lem:accuracy}
To meet a data precision $(\beta,\rho)$,
the privacy budget $\epsilon_s$ for the Laplace mechanism should be lower-bounded by 
\begin{equation}\label{ineq:epsi}
-\frac{\Delta(x)\cdot\ln{(1-\rho)}}{\beta x_{max}}.
\end{equation}
\end{lem}
For the sake of brevity,
we defer the proof to Appendix.
The proofs of later
lemmas/theorems can also be found in this appendix.

Our randomized mechanism \texttt{BR}
is established in the local setting
where the
truthful
mechanisms
are
pushed
to the client side.
The job steps of \texttt{BR} are as follows.
For the given privacy budget $\epsilon_s$ and data sensitivity $\Delta(x)$,
it first calculates a scale parameter $\lambda_i=\frac{\Delta(x)}{\epsilon_s}$.
When a new data
$x_i^{(t)}$ arrives at time $t$,
it draws a random noise $\eta^{(t)}_i$ from the Laplace distribution centered at $0$ with $\lambda_i$,
i.e., $\eta^{(t)}_i\sim {\rm Lap}(0,\lambda_i)$.
Note that the Probability Density
Function (PDF) of this distribution
is 
\begin{equation}\label{eq:laplace}
f_L(o|\lambda_i)=\frac{1}{2\lambda_i}{\rm exp}\big({-\frac{|o|}{\lambda_i}}\big), \quad \lambda_i> 0.
\end{equation}
Then,
it obfuscates the original data $x^{(t)}_i$
in the local device of $u_i$ as $y^{(t)}_i=x^{(t)}_i+\eta^{(t)}_i$.
If $\epsilon_s$ satisfies precision $(\beta,\rho)$,
it directly returns the perturbed value $y^{(t)}_i$.
Otherwise,
it follows the clampdown technology~\cite{mcsherry2009differentially} to
restrict the output to fall inside the data range $[x_{min}, x_{max}]$.
That is,
if an user $u_i$'s noisy data is
$<x_{min}$, its value is set to $x_{min}$; if its value is $>x_{max}$, it is
set to $x_{max}$.  
Pseudocode for \texttt{BR} is presented in Algorithm~\ref{alg:loc_ran},
which implements the above
idea.
All data contributors replace the original data via our proposed randomizer,
and the set of $n$ perturbed values at time $t$
is denoted by $[y^{(t)}]=\{y^{(t)}_1,y^{(t)}_2,\ldots,y^{(t)}_n\}$.
The following theorem shows the privacy guarantee of 
 \texttt{BR}.

\begin{mytheo}\label{theo:BR_gur}
For each data contributor $u_i$ with privacy budget $\epsilon_s$,
our randomized mechanism \texttt{BR} can preserve $\epsilon_s$-LDP.
\end{mytheo}

\begin{algorithm}[t]
\DontPrintSemicolon
\caption{Budget-aware Randomizer \texttt{BR}}
\label{alg:loc_ran}
\begin{algorithmic}[1]
\REQUIRE $x^{(t)}_i$, $\epsilon_s$, $\Delta(x)$, $\beta$, $\rho$ and $[x_{min}, x_{max}]$; \\
\tcp{$x^{(t)}_i$: $u_i$'s personal data}
\tcp{$\epsilon_s$: privacy budget}
\tcp{$\Delta(x)$: neighboring relationship}
\tcp{$\beta$, $\rho$: precision parameters}
\tcp{$[x_{min}, x_{max}]$: data range}
\ENSURE  Perturbed data $y^{(t)}_i$;
\STATE Let $\lambda\leftarrow \frac{\Delta(x)}{\epsilon_s}$;
\STATE Sample random noise $\eta^{(t)}_i$ from  ${\rm Lap}(0,\lambda_i)$;
\STATE Let $y^{(t)}_i \leftarrow x^{(t)}_i+ \eta^{(t)}_i$;
\IF{$\epsilon_s\geq-\frac{\Delta(x)\cdot\ln{(1-\rho)}} {\beta x_{max}}$}
    \RETURN $y^{(t)}_i$; 
    \STATE ~\tcp{Upper bound for clamps}
    \IF{$y^{(t)}_i\!>\!x_{max}$} 
    \RETURN $x_{max}$; 
    \STATE \ \tcp{Lower bound for clamps}
    \IF{$y^{(t)}_i\!<\!x_{min}$ }
    \RETURN $x_{min}$;
    \ELSE
    \RETURN $y^{(t)}_i$;
    \ENDIF
    \ENDIF
\ENDIF
\end{algorithmic}
\end{algorithm}

\subsection{Design of Shuffler}\label{subsec:shuf}
Next,
we introduce a random shuffler \texttt{RS}
to achieve sender anonymity.
Our design is built upon the Mallows mechanism,
but differs in two fundamental aspects:
(\romannumeral 1)
we focus on
how to come up with
a good partition,
instead of using fixed-size groups,
and (\romannumeral 2)
we introduce an auxiliary technology to improve the robustness,
ensuring that our shuffler performs effectively 
even in the most extreme cases.

\noindent{\bf Data Permutation.}
Recall that
there are $n$ data contributors,
each outputting a noisy response $y_i$.
Like the shuffle model in~\cite{Balle2019Privacy, Albert2019Distributed},
we have an external shuffler
to receive the outputs
and sends a permutation of them to the next destination.
In particular,
the shuffler can be implemented through a protocol run by each data contributor,
e.g. using mix-net~\cite{shen2021daenet} or doubly-encrypted technology~\cite{Bittau2017Prochlo}.
Assume that data arrival at the shuffler
is captured by a \emph{first-come-first-serve} discipline.
Commonly,
we will use the
following notations:
$\vec{y}$ denotes a data sequence of $[y]$;
${\rm Idx}(y_i)=j$ means that $y_i$ is at the index $j$ in $\vec{y}$.
As an example,
for sequence $\vec{y}=<y_4, y_3 ,y_1, y_2>$,
we say that the noisy response $y_3$ is at index $2$,
i.e., ${\rm Idx}(y_3)=2$.
Then,
$\sigma(\vec{y})$ denotes
rearranging the sequence $\vec{y}$ according to
a permutation $\sigma\in S_n$.

It can be argued that
the index of $n$ elements in $\vec{y}$
is a special permutation.
The \emph{normalized form } of the permutation is represented
in the classic one-line notation:
\begin{equation}\label{eq:data_perm}
\sigma_{\vec{y}}=\big(\,{\rm Idx}(y_1), \ {\rm Idx}(y_2),\ \ldots \  ,{\rm Idx}(y_n)\,\big).
\end{equation}
We refer to Eq.~(\ref{eq:data_perm}) as
a data permutation.
As mentioned above,
the all data contributors are partitioned into $k$ non-overlapping groups $\{G_1,\ldots, G_{k}\}$,
based on which every permutation $\sigma\in S_n$
can be consider as the union of $k$ groups as well.
We abuse the same symbol to denote the groups on permutations,
and
call this an initial partition $\xi_0$.
Next,
the \emph{group form} of permutations under $\xi_0$ is
represented by applying function $\boldsymbol{\sigma}(\cdot)$ to each group, i.e.,
$\mathcal{G}_{\xi_0}({\sigma})=\,\boldsymbol{\sigma}(G_1), \ \boldsymbol{\sigma}(G_2),\ \ldots \  , \boldsymbol{\sigma}(G_k)$.
For example,
when $\sigma_{\vec{y}}=(1 ,2 ,4 ,3, 5)$,
$\mathcal{G}_{\xi_0}(\sigma_{\vec{y}})$ can be $\big( 4, 1, 3\big),\big( 2, 5\big)$.
For each permutation $\sigma\in S_n$,
we could define a vector
$\delta^{\sigma}=(\delta^{\sigma}_1, \ldots \delta_{k}^{\sigma})$,
where $\delta^{\sigma}_l$ $ (l=1,\ldots, k)$ is the number of elements in group $G_l$.
For instance,
the vector of $\sigma_{\vec{y}}$ is $(3, 2)$ in the above example.
Assuming $\sigma$ and $\sigma'$ are two permutation in $S_n$.
When $\mathcal{G}_{\xi_0} (\sigma)\neq \mathcal{G}_{\xi_0}\mathcal(\sigma')$ and for all $\delta_l$,
$\forall l\in \{1,\ldots,k\}$,
we think that $\sigma$ and $\sigma'$ are neighboring under a given partition $\xi_0$,
i.e., $\sigma \simeq_{\xi_0} \sigma'$.
Hence,
the set of all neighboring permutations can be expressed as 
$N_{\xi_0}=\{(\sigma, \sigma')\mid \sigma \simeq_{\xi_0} \sigma'; \sigma, \sigma'\in S_n\}$.

\begin{mydef}[Group width~\cite{Meehan2021Privacy}]\label{def:sensitivity}
Given a permutation $\sigma$
and a partition $\xi_0$,
let $d(\cdot, \cdot)$ be the distance measure,
the group width measures
the maximum change between the members of a group on $\sigma$,
i.e.,
\begin{equation}
\mathcal{W}(\sigma,\, d,\,G_l)={\max}_{i, j\in G_l}|\boldsymbol{\sigma}^{-1}(i)-\boldsymbol{\sigma}^{-1}(j)|, \ \ \forall G_l\in\xi_0. \notag
\end{equation}
\end{mydef}
Obviously,
the resulting width is a function of parameters $\sigma,\, d$ and $G_l$.
Then we denote the global width as
\begin{equation}
    \omega_{\xi_0}={\max}_{G_l\in \xi_0}\mathcal{W}(\sigma,\, d,\,G_l).
\end{equation}
The magnitude of randomness added by Mallows mechanism
depends on sensitivity,
which is the largest possible difference over all neighboring permutations in $N_{\xi_0}$.
If the distance metric
is the Kendall tau distance $d_K(\cdot,\cdot)$,
the sensitivity of the central permutation $\sigma_0$
can be given by
\begin{equation}\label{eq:kd_sens}
\Delta(\sigma_0:\, d_K,\, \xi_0)=\frac{\omega_{\xi_0}(\omega_{\xi_0}+1)}{2}.
\end{equation}
The main rationale behind Eq.~(\ref{eq:kd_sens}) can be
found in~\cite{Meehan2021Privacy},
using the similar principle as bubble sort.
To achieve a better performance,
the Mallows mechanism
optimizes the central permutation $\sigma_0$ for reducing
sensitivity over distance measure.
The privacy property of the mechanism is as follows.

\begin{lem}[Meehan et al.~\cite{Meehan2021Privacy}]\label{lem:mallows}
     Given the group partition $\xi_0$ and 
     the privacy parameter $\alpha$,
     the Mallows mechanism provides $(\alpha, \xi_0)$-$d_{\sigma}$ privacy.
\end{lem}

Unfortunately,
this scheme is sub-optimal
because,
as shown in the examples
below,
it does not take the impact of partitions on the performance into consideration.

\begin{myexam}\label{exa:per_1}
Consider a 6-element central permutation $\sigma_0=(1,4,5,6,3,2)$.
Suppose that all users are divided
into two disjoint groups $G_1$ and $G_2$.
Assume that $\boldsymbol{\sigma}(G_1)=(1,4,5,6,3)$ and $\boldsymbol{\sigma}(G_2)=(2)$.
By Eq.~(\ref{eq:kd_sens}),
the sensitivity of Mallows mechanism over $d_K$ is $15$.
If we rectify the user groups into $G_1=\{1,4,5\}$ and $G_2=\{6,3,2\}$,
the sensitivity of the refined partition will significantly decrease to $10$.
\end{myexam}

\begin{myexam}\label{exa:per_2}
We also consider two ``extreme'' cases.
At one extreme,
each group contains exactly one data contributor (i.e.,
when $k = n$, where $k$ and $n$ are the number of groups and data contributors,
respectively).
In this case,
the sensitivity of the Mallows mechanism is equal to $0$,
which takes the shuffler out of service due to $\theta\rightarrow \infty$.
At the opposite extreme,
the set of $n$ data contributors is considered as a single partition (i.e., $k = 1$).
With the increasing size of $n$,
the Mallows model becomes more similar 
to the uniform distribution,
for which an awful lot of low-complexity algorithms 
can be applicable to generate permutations.
\end{myexam}

These two examples corroborate that
(\romannumeral 1) the Mallows mechanism is ideal because
it represents the best-case scenarios,
and (\romannumeral 2) there exist fertile grounds to adjust
it according to different partition schemes.
By applying a two-case analysis,
we
aim to
design a
more sophisticated shuffler tailored to
the aforementioned problems.
To do so,
we must know
the exact value of shuffle degree $\alpha$ and the number of groups $k$.
Without
loss of generality,
assume that $\alpha$ and $k$ are the input parameters
in our implementation.
Then,
we can improve the Mallows mechanism in the two cases.

\noindent{\bf When Partition Is Unequal:}
This situation coincides with Example~\ref{exa:per_1},
for which we concentrate on how to
generate an appropriate partition.
Following the same general way as that of the Mallows mechanism,
our design should strike an intuitive balance
between utility and privacy.
This objective necessitates
the need of defining a target function,
the
optimization of which leads to
the decreasing of the privacy budget.
To this end,
we formulate the Grouping Refinement Problem (GRP),
with the goal of
minimizing the sensitivity $\Delta(\sigma_0:\, d_K,\, \xi)$ as follows.

\begin{mydef}[Grouping refinement problem]\label{def:refine}
Given a central permutation $\sigma_0$,
a distance measure $d(\cdot,\cdot)$,
and a set of data contributors $u_1$, $\ldots$, $u_n$,
partition the $n$ data contributors into $k$ $(1<k<n)$ groups $\xi=\{G_1, \ldots, G_k\}$
such that
the global sensitivity $\Delta(\sigma_0:\, d,\, \xi)$
is minimized,
subject to $G_l \cap G_m =\emptyset$ for any $l\neq m$.
\end{mydef}
In the definition,
the difficulty is that
to find the minimum value of sensitivity,
every possible assignment between $n$ data contributors and $k$ disjoint groups should be derived,
which involves amounts of computation.
We show the complexity of the problem by the following theorem.

\begin{mytheo}\label{theo:Np}
The GRP problem is NP-hard.
\end{mytheo}

Given the
hardness of this problem,
we resort to heuristic algorithms.
If we cluster the members of each group closely together,
a lower width (or sensitivity) can be achieved.
Specifically,
we consider using the agglomerative clustering
method.
At the initial phase,
each data contributor forms a singleton group.
Then,
we iteratively choose and merge two groups that result
in the minimum value increase of the objective function
among all group pairs.
This iterative process continues to
run until there exist $k$ groups. 
Through this scheme,
the data contributors who are close in connectivity
could be gradually clustered together.
In the same way as~\cite{Meehan2021Privacy},
we next (\romannumeral 1) compute the dispersion parameter $\theta=\frac{\alpha}{\Delta(\sigma_0)}$,
(\romannumeral 2) sample a permutation $\hat{\sigma}$ from the Mallows model ${\rm Pr}\big{[}\sigma\mid \sigma_0, \theta\big{]}$,
and (\romannumeral 3) apply the desired permutation for shuffling.

\begin{figure}
  \centering
  \setlength{\belowcaptionskip}{-0.1cm} 
  \includegraphics[width=.43\textwidth]{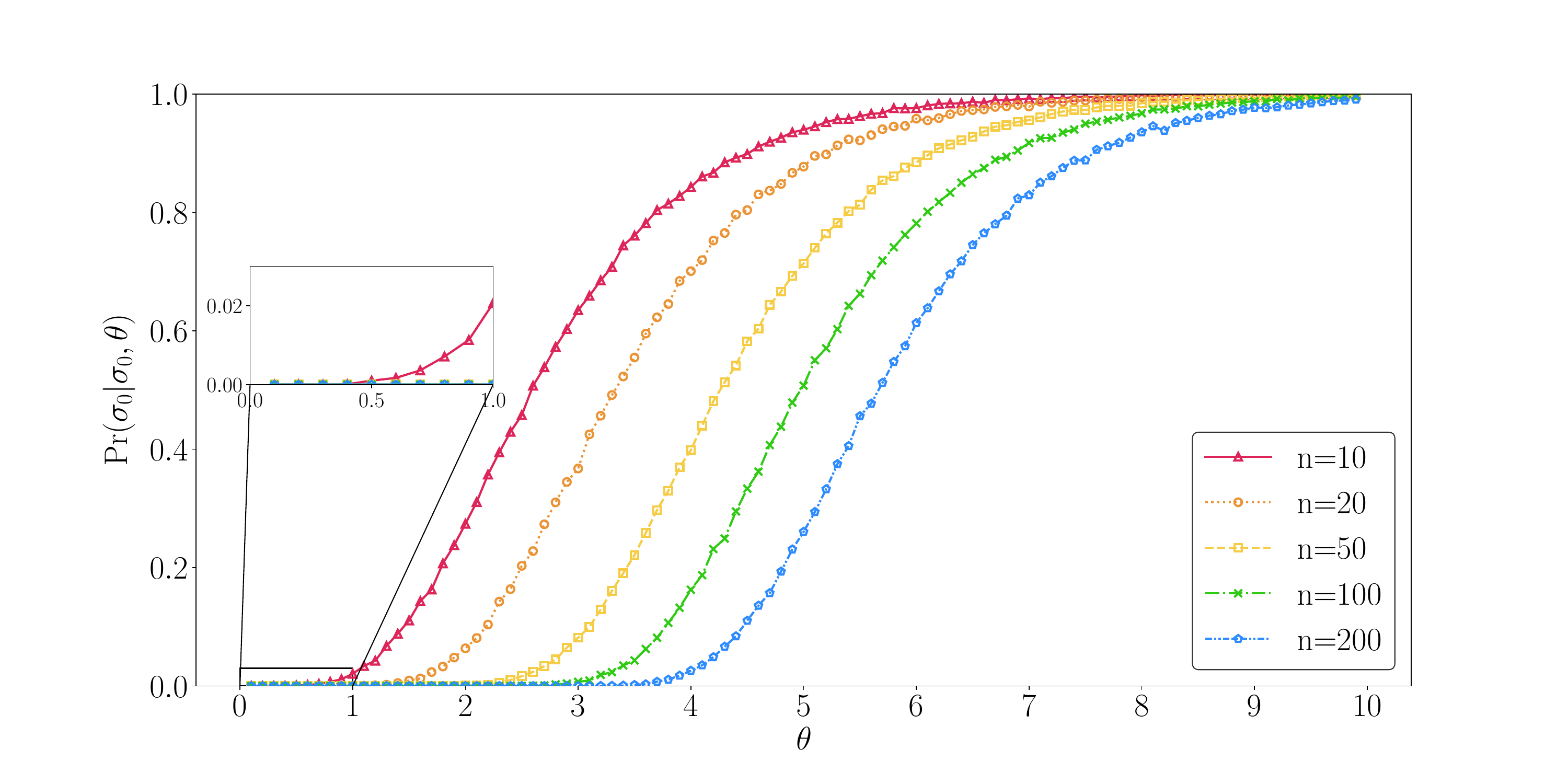}
  \caption{
The probability assigned to central permutation $\sigma_0$
for different parameters $\theta$ and $n$.}\label{fig:Mallows_equal}
\end{figure}

\noindent{\bf When Mallows Mechanism Fails:}
From Example~\ref{exa:per_2},
we can see that
if $k = n$,
the shuffler of the Mallows mechanism does not do anything.
This means that
only the local randomizer is available due to ${\rm Pr}\big{[}\sigma_0 \mid \sigma_0, \infty\big{]}=1$,
which can be observed from Fig.~\ref{fig:Mallows_equal}.
On the other hand,
when $n = 1$,
the shuffler also suffers from a non-negligible performance degradation.
The reason accounts for the failure
is that 
a large sensitivity $\Delta(\sigma_0)$ worsens the privacy budget $\alpha$ while
the distribution of the Mallows model 
gets closer to the uniform distribution 
(this can be observed from Fig.~\ref{fig:Mallows}).
To overcome these obstacles,
we must answer the question:
how do we define one criterion for formalizing
the capability boundaries of the Mallows mechanism.
Practically,
$\theta$ is chosen to be a relatively small value~\cite{li2020extended}, e.g., between $0$ and $1$.
We pick $0.1$ as the lower bound of $\theta$,
and thus
$\Delta(\sigma_0)$ can be limited to
an approximated closed range of $[\alpha, 10\alpha]$.

When the Mallows mechanism is no longer sufficient for obtaining  ``good'' results,
we introduce
an efficient in-place algorithm,
analogously to the Fisher-Yates shuffler~\cite{knuth2014art},
to uniformly generate a random permutation.
It starts with the central permutation $\sigma_0$
and includes $n-1$ steps.
At the $i$th step,
it chooses a random position $j$ uniformly from $\llbracket i+1,n \rrbracket$,
and swaps the current term in $j$ with the item at position $i$.
Note that the auxiliary mechanism protects identity privacy against adversary's rebuilding based on a series of random interchange operations.
We provide below the risk analysis for  
the auxiliary shuffler.


\begin{lem}\label{lem:shuff}
For an $n$-dimensional permutation $\sigma\in S_n$,
the auxiliary shuffling mechanism
guarantees that 
the probability of an adversary identifying $\sigma$ is $\frac{1}{n-1}!$.
\end{lem}

To sum up,
we design a
more robust shuffling mechanism,
whose
details are given in Algorithm~\ref{alg:shuffler}. 
One could think that our modification could destroy the privacy guarantee of the Mallows mechanism. 
We use the following theorem to reveal that this is not a case.

\begin{mytheo}\label{theo:d_privacy}
    Given the privacy parameter $\alpha$,
    our shuffle mechanism \texttt{RS} still preserves $(\alpha, \xi)$-$d_{\sigma}$ privacy.
\end{mytheo}

\begin{algorithm}[t]
\caption{Robust Shuffler~\texttt{RS}}
\label{alg:shuffler}
\begin{algorithmic}[1]
\REQUIRE $\vec{y}^{(t)}$, $k$, $G=(V,E)$ and $\alpha$; \\
\tcp{$\vec{y}^{(t)}$: noisy input sequence at time $t$}
\tcp{$k$: number of groups}
\tcp{$G=(V,E)$: relationship graph}
\tcp{$\alpha$: privacy budget}
\ENSURE Anonymized dataset $[z^{(t)}]$;
\STATE Calculate data permutation $\sigma_{\vec{y}^{(t)}}$ according to Eq.~(\ref{eq:data_perm});
\STATE Extract initial partition $\xi_0$ from $G$;
\STATE Calculate sensitivity $\Delta(\sigma_0)$ according to Eq.~(\ref{eq:kd_sens});
\IF {$\alpha \leq  \Delta(\sigma_0) \leq 10\alpha$} 
\STATE $\xi\leftarrow GroupRefinement(\sigma_{\vec{y}^{(t)}},k)$; 
\STATE Recalculate $\Delta(\sigma_0)$ according to Eq.~(\ref{eq:kd_sens});
\STATE $\theta \leftarrow \frac{\alpha}{\Delta(\sigma_0)}$;
\STATE Sample a random permutation $\tau$ from the Mallows model with center $\sigma_{\vec{y}^{(t)}}$ and spread parameter $\theta$;
\STATE $\sigma^* \leftarrow \sigma_{\vec{y}^{(t)}}^{-1}\tau$; 
\ELSE
\STATE $\sigma^* \leftarrow \sigma_{\vec{y}^{(t)}}$;
\FOR {$i\leftarrow 1, 2, \ldots, n-1$}
\STATE Choose at random a integer $j$ in $\llbracket i+1,n \rrbracket$;
\STATE Interchange $\boldsymbol{\sigma}^*(i)$ and $\boldsymbol{\sigma}^*(j)$;
\ENDFOR
\ENDIF
\STATE  $[z^{(t)}]\leftarrow \{y^{(t)}_{\boldsymbol{\sigma}^*(1)}, y^{(t)}_{\boldsymbol{\sigma}^*(2)}, \ldots, y^{(t)}_{\boldsymbol{\sigma}^*(n)}\}$;
\RETURN $[z^{(t)}]$;
\end{algorithmic}
\end{algorithm}

\subsection{Design of Estimator}\label{subsec:estm}
The estimator \texttt{ET} is a mathematical rule used to create
the mean estimate based on the data
that are affected by the randomness imposed due to the previous modules.
Let $[z]=\{z_1, z_2, \ldots, z_n \}$ be the  output private data from the shuffler
and $\hat{\Theta}$ be an estimation of the mean of $[z]$. 
We apply several alternative estimation procedures to generate $\hat{\Theta}$.
 
\begin{enumerate}[0]
    \item[$\bullet$]  The \underline{\textit{sample estimator}} is the standard method for the average computation in data aggregation,
    and returns the average of the noisy data as the estimate of the aggregated data, i.e.,
    \begin{equation}
     \hat{\Theta}_{SE}=\frac{1}{n}{\sum}_{i\in\llbracket 1,n \rrbracket} z_i ~.
    \end{equation}
    \item[$\bullet$]  The \underline{\textit{maximum-likelihood estimator}} returns the solution of the likelihood function as the estimate of the aggregated data.
    Based on the Laplacian PDF,
    we have $z_i\sim {\rm Lap}(\hat{\mu},\lambda_i)$,
    where $\hat{\mu}$ is the unknown parameter (mean) to be estimated.
    We formulate the likelihood function as
    \begin{equation}
        \ell(\hat{\mu};[z])=\prod_{i=1}^n \frac{1}{2\lambda_i}{\rm exp}({-\frac{|z_i-\hat{\mu}|}{\lambda_i}}) ~.
    \end{equation}
    Then, its logarithmic form can be written as
    \begin{equation}
        \log\ell(\hat{\mu};[z])=-n\log(2\lambda_i)-\frac{1}{\lambda_i}{\sum}_{i\in \llbracket 1,n \rrbracket} |z_i-\hat{\mu}|~.
    \end{equation}
    Given default parameters $n$ and $\lambda_i$,
    the maximum likelihood estimator is 
    \begin{equation}
\hat{\Theta}_{MLE}=\arg\min{\sum}_{i\in \llbracket 1,n \rrbracket} |z_i-\hat{\mu}|~.
    \end{equation}
    \item[$\bullet$]  The \underline{\textit{bootstrap   estimator}}~\cite{tibshirani1993introduction} returns an estimation by randomly selecting the sample data and 
    then performing inference over 
    the newly constructed resamples (bootstrap samples). 
    Suppose that the number of the bootstrap samples is indicated by $B$. 
    We denote the $b$th bootstrap sample by
    $[\tilde{z}^b]\!=\!\{\tilde{z}^b_1,\tilde{z}^b_2,\ldots, \tilde{z}_n^b \}$
    and draw it uniformly according to 

    \begin{equation}
        \tilde{z}^b_i\overset{{\rm i.i.d.}}{\sim} {\rm Unif}\big([z]\big), ~~i=1,2,\ldots,n~.
    \end{equation}
 Taking the average of any bootstrap sample, 
 we have:
    \begin{equation}
        {\rm Avg}([\tilde{z}^b])=\frac{1}{n}{\sum}_{i\in \llbracket 1,n \rrbracket} \tilde{z}^b_i~.
    \end{equation}
 The bootstrap estimator will then be:
    \begin{equation}
       \hat{\Theta}_{BS}=\frac{1}{B}{\sum}_{b\in \llbracket 1,B \rrbracket} {\rm Avg}([\tilde{z}^b])~.
    \end{equation}
\end{enumerate}

To better understand the impact of the adopted estimators on noise smoothing,
we give an example illustrating what they finally produce.
Moreover,
we will empirically evaluate 
these estimators in Section~\ref{sec:Evaluation}.

\begin{myexam}\label{exa:est_outflier}
Suppose we have an array of raw IoT data $[x^*]=\{4, 2, 1, 3, 5\}$ and the corresponding noisy outputs $[z^*]=\{9.5, 1.1, 8.4, 2.8, 3.2\}$ from the shuffler.
Clearly,
the true average of $[x^*]$ is equal to $3$
and two extreme values differ from most values in $[z^*]$.
Let us see what happens to the mean when the  three alternatives are used.  
The sample mean will be $\hat{\Theta}_{SE}=5$,
which moves away the true value by about $67\%$ and is thus being overestimated.
The maximum-likelihood method picks $\hat{\Theta}_{MLE}=3.2$ as the estimation average and is more
efficient than the former.
Assume that the bootstrap estimator resamples $2$ times from $[z^*]$, 
and gets $\{1.1,2.8,1.1,3.2,2.8\}$
and $\{9.5,2.8,3.2,3.2,1.1\}$.
On the same scale,
the bootstrap estimator performs the best because $\hat{\Theta}_{BS}=3.08$.
\end{myexam}




\subsection{Putting Things Together}\label{subsec:Put_Thing}
Now,
we can integrate the previous modules
into the full \pmb{\texttt{RASE}} framework,
which consists of three basic parts,
connecting together (\romannumeral 1) 
the local perturbation mechanism \texttt{BR},
(\romannumeral 2) the anonymization  mechanism \texttt{RS} and (\romannumeral 3)
the mean estimation mechanism  \texttt{ET}.
The overall \pmb{\texttt{RASE}} works as follows.
In the initial stage,
the fusion center disseminates a message 
to all data contributors within the system.
This message outlines the specified precision $(\beta, \rho)$ that requires compliance.
At each time $t$,
the data contributor runs \texttt{BR} to her own data $x_i^{(t)}$
and subsequently passes the masked data $y_i^{(t)}$ to a mediation component.
Upon collecting data from all sources,
the mediation component employs \texttt{RS} to make the sender anonymous and presents the confidential dataset $[z^{(t)}]$ to the fusion center.
The fusion center invoke \texttt{ET} to yield an approximate answers, aimed at fulfilling the request made by subscribers.
Algorithm~\ref{alg:P3DA} spells out the whole steps. 

\begin{algorithm}[t]
\caption{\pmb{\texttt{RASE}}: Randomize-Shuffle-Estimate based Data Aggregation}
\label{alg:P3DA}
\begin{algorithmic}[1]
\REQUIRE  $\{u_i\mid i\in \llbracket n \rrbracket\}$ and $(\beta,\rho)$; \\
\tcp{$\{u_i\mid i\in \llbracket n \rrbracket\}$: set of $n$ data contributors}
\tcp{$\beta,\rho$: precision requirement}
\ENSURE Estimated average $\hat{\Theta}$;
\STATE Fusion center broadcasts requirement $(\beta,\rho)$;\\
\tcp{Data contributor-side}
\FOR {$u_1,\ldots,u_n$}
\STATE Apply \texttt{BR} to her data;
\STATE Submit the perturbed data to the shuffler;
\ENDFOR
\STATE Shuffler applies~\texttt{RS} to rearrange the data sequence and then uploads the new sequence to the fusion center;
\tcp{Fusion center-side}
\STATE Fusion center invokes \texttt{ET} to obtain average value $\hat{\Theta}$;
\RETURN $\hat{\Theta}$;
\end{algorithmic}
\end{algorithm}

\begin{table*}[htbp]
\centering  
\caption{Statistics of the data drawn from 
 REFIT}
 \label{tab:dataset}  
\begin{tabular}{|c|c|c|c|c|c|}
\hline
   & & & & &\\[-8pt]  
   &ID &  Device Types & $\#$ of Sensors& Data Range& $\#$ of Months Used  \\ \hline
   & & & & & \\[-6pt]  
   {\bf Training Houses} &  House 1-House 17 & $24$& $\geq 11$& $[3.900,178.300]$ & $8$\\ \hline
   & & & & &\\[-6pt]  
   {\bf Test Houses} & House 18-House 21 & $11$& $\geq 13$ & $[11.800,99.027]$& $14$\\ \hline
\end{tabular}
\end{table*}

\noindent \textbf{Utility Analysis.} 
Since the proposed mechanisms 
leverage randomness to mitigate the privacy threats, 
it is essential to assess the potential impact on the data utility.
The following theorem helps for establishing the utility guarantee of \pmb{\texttt{RASE}}.

\begin{mytheo}\label{theo:MAE}
    When uses the empirical mean as the estimator,
    the absolute expected error $\mathbbm{E}\big\|\hat{\Theta}_{SE}- \Theta\big\|_1$ is bounded by $\max\big(O(\frac{1}{\epsilon_s}),\Delta(x)\big)$.
\end{mytheo}

While we only examine the theoretical performance for the mean estimator,
the empirical comparisons of the three statistical techniques are conduced in Section~\ref{sec:Evaluation}.
The results reveal that the maximum-likelihood estimator often produces the  best estimate of observing the data.

\noindent \textbf{Privacy Analysis.} 
The privacy guarantee provided by our \pmb{\texttt{RASE}} paradigm
can be easily deduced from Theorems \ref{theo:BR_gur} and \ref{theo:d_privacy}; it just uses the post-processing of LDP.
Here,
we further discuss how \pmb{\texttt{RASE}} safeguards against attacks from both the fusion center and subscribers during the actual operation. 

\subsubsection{Fusion center's attacks}
As we mentioned in Section~\ref{subsec:threat},
the fusion center has two
different methods to probe into
data contributor's privacy:
(\romannumeral 1)
recovering original data
with some prior knowledge,
such as
the global data distribution,
and (\romannumeral 2)
attempting to re-identify where the data come from.

For the first method, our discussion revolves around the following process.
After implementing personalized noise and random shuffling,
the noisy data $y_i$,
instead of $x_i$,
is submitted to the fusion center.
As a result,
the fusion center's view is limited
to the masked dataset $[y]$.
The inherent randomness obscures the private data of each data contributor $u_i$, 
rendering it intricate to infer the original value of $u_i$'s data through observing the imprecise output.
This ensures that the data obtains a substantial privacy guarantee via the one-shot perturbation.  



In the case of the second method,
we describe
a novel, data-independent shuffling technology
to prevent data from being re-identified.
Note that
the shuffler receives data from multiple sources, rearranges them, and then transmits them in a randomized sequence to the next destination 
as well as
the fusion center.
This results in the adversary perceiving a uniformly random permutation of the received data.
In order to ascertain the sender's identity,
an adversary has to
identify the target data inside an unorder set.
This is hard to achieve when the set contains vast amounts of elements.
\subsubsection{Subscriber’s attacks}
Under natural constraints,
this type of adversary is dramatically weaker than
the previous one, but possesses the capability to conduct eavesdropping attacks.
Protecting
the security of communication contents is an ``orthogonal'' problem to our study.
Fortunately,
it can be
solved via encryption-based solutions.
We can take the existing solutions (e.g., \cite{mohammed2016efficient} and \cite{wang2018secure}) as supplements to enhance our \pmb{\texttt{RASE}} paradigm.
In this fashion,
only encrypted
data can be intercepted by unscrupulous subscribers,
thus breaking the cryptosystem without the key a formidable challenge.

\noindent \textbf{Complexity.} 
Next, we delve into the time complexity of the \pmb{\texttt{RASE}} paradigm,
according to how it carries out the privacy-process steps.
On the data contributor side,
the time consumed depends on the process of perturbing the data using the local randomizer \texttt{BR}.
In practice, \texttt{BR} generates a random noise from Laplace distribution
and adds it to the true data,
which can be done in a time of $O(1)$.
For the shuffler $\texttt{RS}$,
it is easy to verify that with the size of the input data, the time cost grows polynomially,
taking $O(n^3)$ time.
On the fusion center side,
the time complexity varies with the adopted 
estimator.
Assuming that the fusion center needs $\Psi_{ES}$ time to obtain the mean estimate, the overall time complexity
of \pmb{\texttt{RASE}} is $\Psi_{ES}+O(n^3)$. 
\section{Experimental Evaluation}
\label{sec:Evaluation}
In this section, we compare \pmb{\texttt{RASE}} with
the baseline solution \texttt{BR}
and two other schemes proposed in the relevant literature.


\subsection{Experimental Setup}
The construction of the data contributor's privacy specification is as follows.
We chose 
the value of $\epsilon_s$ 
from $[0.01, 10]$.
Another privacy budget $\alpha$ is set at $30$ as default.
The confidence
parameters $\rho$ and $\beta$ are fixed at $0.9$ and $0.5$, 
respectively.
In order to implement \pmb{\texttt{RASE}} in a Python environment,
we use several open-source third-party libraries including Pandas, Matplotlib and Sklearn.
In each experiment, 
we vary one parameter while keeping the others constant. 
Besides,
all evaluation results are the average of
$10$ trials.
These operarions facilitate the observation of how that particular factor influences the performance.

To the best of our knowledge, there is no existing algorithms that can directly tackle the problem  discussed in this paper.
We analyze three fundamental methods 
in an attempt to
adapt them to our setting:
(\romannumeral1) in \texttt{BR}, each data contributor generates independent random noise using Algorithm~\ref{alg:loc_ran}, and then perturbs the data to be reported (no shuffling);
(\romannumeral2) the Laplace-fisher algorithm
performs local randomization by using the Laplace mechanism~\cite{Dwork2013The},
and then runs the Fisher-Yates~\cite{knuth2014art} shuffler
to reorder the data sequence;
and (\romannumeral3) \texttt{BR}-mallows
applies the \texttt{BR} algorithm to 
obfuscate the true data,
but a permutation sampled from the Mallows distribution is used to shuffle the data sequence.

\noindent{\bf Datasets.}
The source data are obtained from the 
REFIT smart home dataset~\cite{Firth2017REFIT},
which comprises electrical load measurements from $180$ appliances belonging to $21$ households in the U.K.
The data collection spanned from Oct. $2013$\,--\,Jun. $2015$,  
during which the active power consumption (in Watts) was recorded at an $8$-second interval.
Instead of processing the entire dataset,
we randomly select $5$ houses,
and extract the data from the sensors placed within the houses for the evaluation of the time-series data. 
An
overview of the resulting data is shown in Table~\ref{tab:dataset}.


\begin{figure*}[htbp]
\centering
\subfloat[]{\includegraphics[width=.19\linewidth]{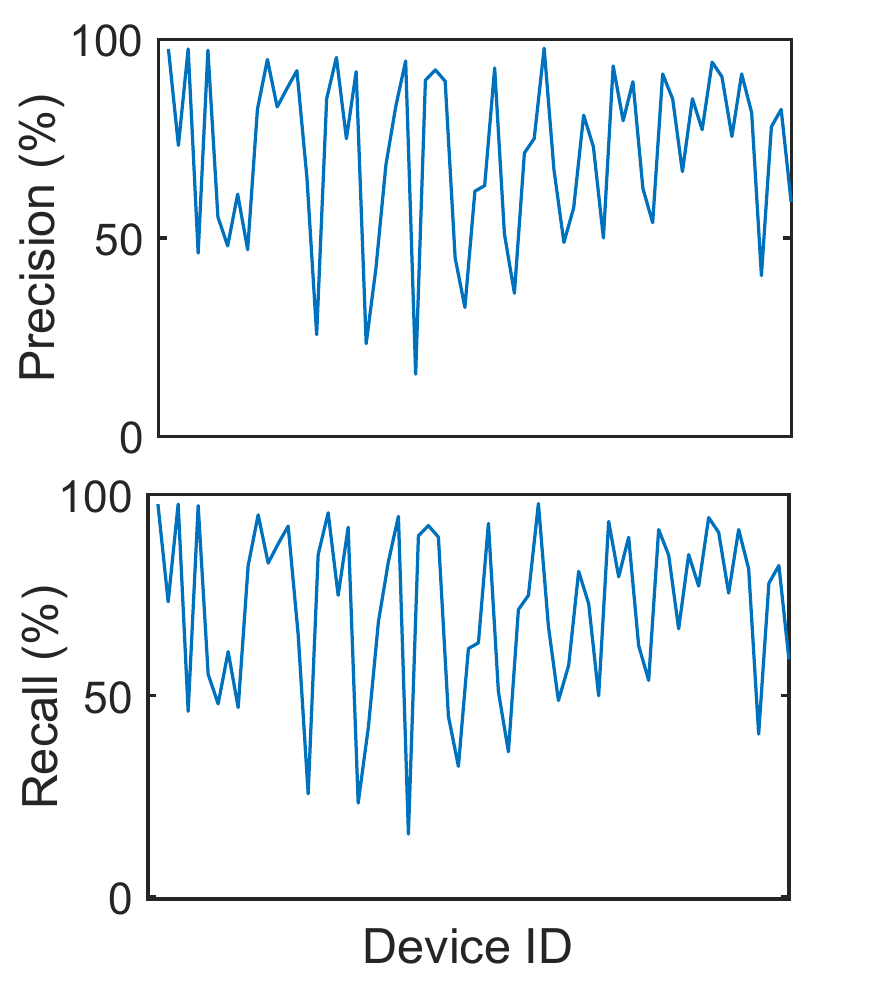}}\hspace{3pt}
\subfloat[]{\includegraphics[width=.19\linewidth]{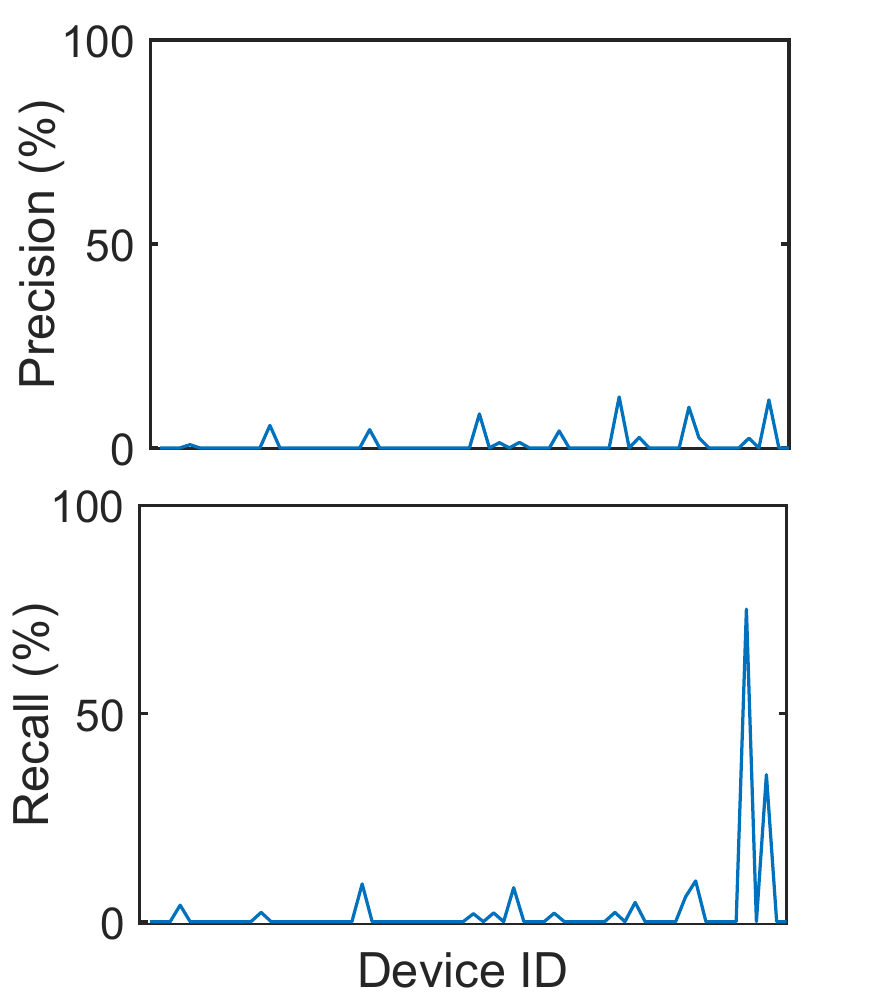}}\hspace{3pt}
\subfloat[]{\includegraphics[width=.19\linewidth]{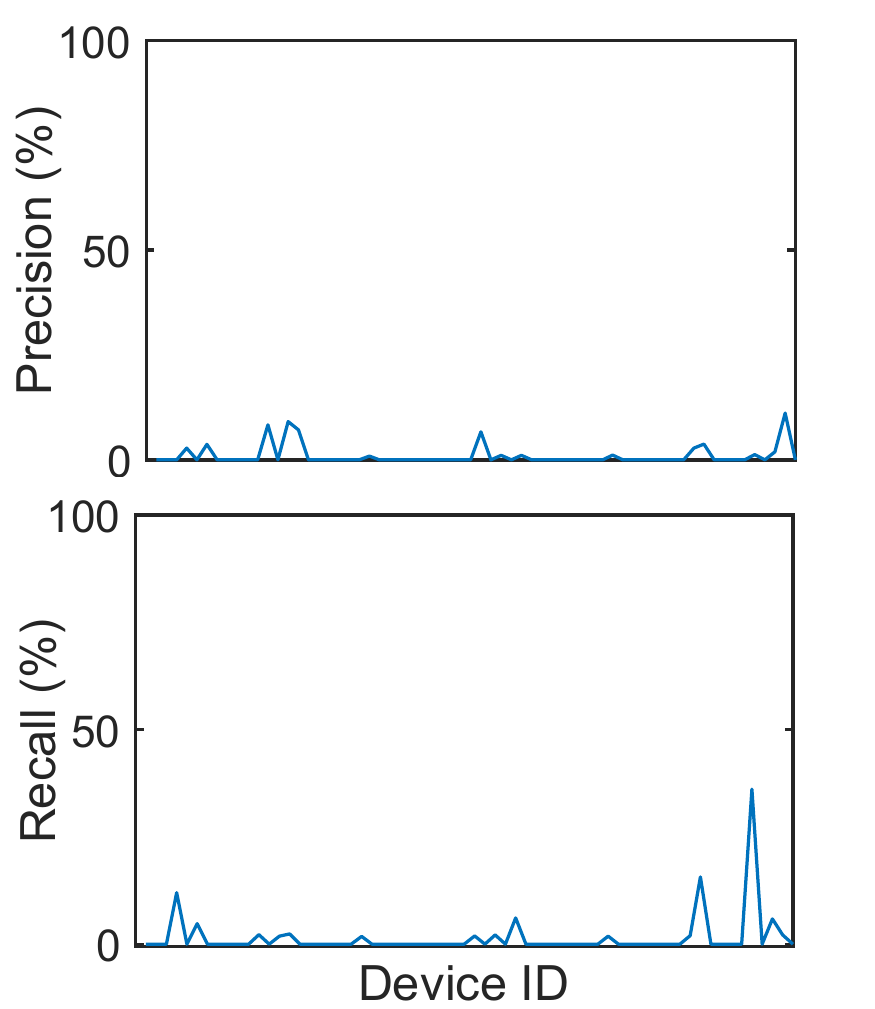}}\hspace{3pt}
\subfloat[]{\includegraphics[width=.19\linewidth]{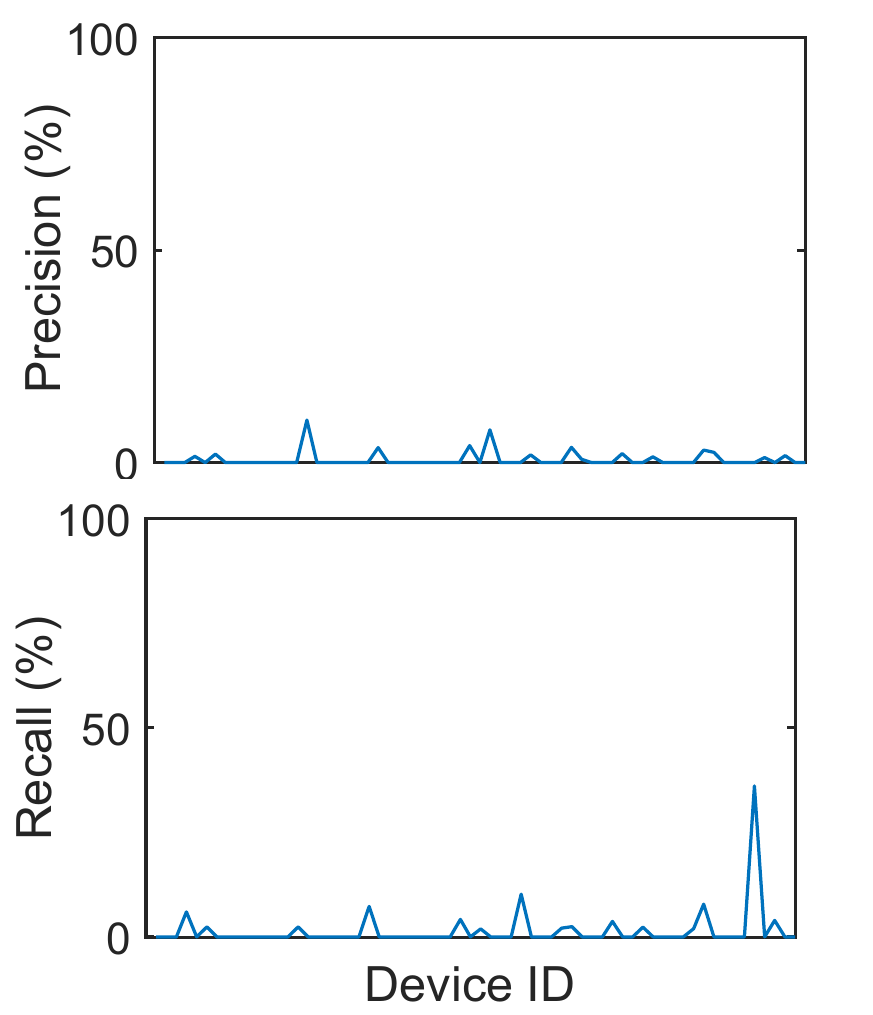}}\hspace{3pt}
\subfloat[]{\includegraphics[width=.19\linewidth]{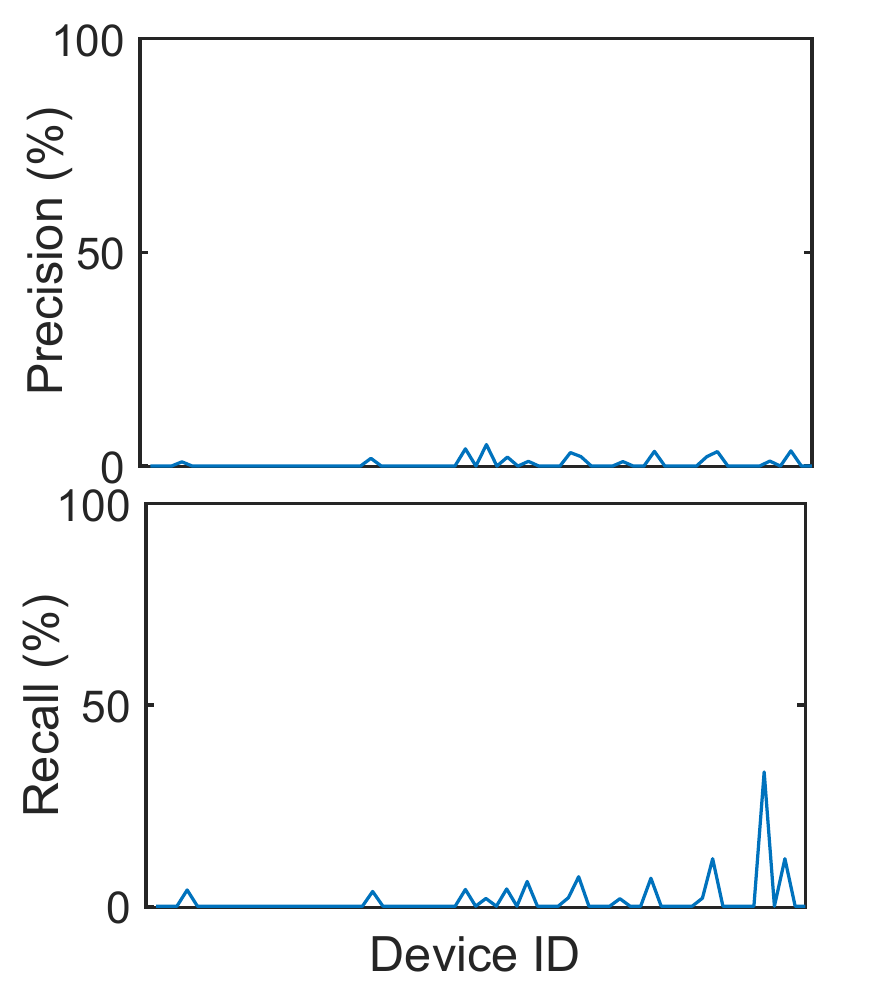}}\\
\caption{Comparison of sanitizing algorithms for $n=64$ sensor devices: (a) precision-recall on the raw data; (b) precision-recall after  \texttt{BR}; (c) precision-recall after  \texttt{BR}-Mallows; (d) precision-recall after Laplace-fisher; (e) precision-recall after  \pmb{\texttt{RASE}}.}\label{fig:privacy_precision}
\end{figure*}

\noindent{\bf Attack Methods.}
Normally,
the adversary
can grab
the features of each user
from the non-anonymized history data through some machine-learning approaches.
Without loss of generality,
we assume that the knowledge of data distribution 
about a specified user
is public.
As a result,
the process of re-identifying data senders
can be considered as the multi-label classification problem.
We adopt the well-established Random Forest algorithm~\cite{yan2020time} 
to train the attack model.
In detail, 
the mentioned data is divided into 
explicit
two groups with the ratio of $4:1$.
The larger one is used for training the classification model, and the other set is used for testing.

\subsection{Resilience of Solutions against  Disclosure Attacks}
We have evaluated the effectiveness of the sanitizing algorithms\,--\,\pmb{\texttt{RASE}},
\texttt{BR}, Laplace-fisher and \texttt{BR}-mallows\,--\,in detecting disclosure attacks based on
the precision-recall metric.

By definition,
precision corresponds to the ability to accurately identify predicted results.
In the context of this work, precision measures how effectively an algorithm can detect attacks instances of attacks. 
Recall, on the other hand, corresponds to the ability to effectively retrieve all relevant instances of a particular class. 
It provides a measure of how well an algorithm can identify the proportion of actual attacks.
The two metrics are then defined as:
\begin{equation}
    \begin{split}
        {\rm Precision}\!=\!\frac{TP}{TP\!+\!FP} \!\times\! 100\%,\,
        {\rm Recall}\!=\!\frac{TP}{TP\!+\!FN}\!\times\! 100\%,
    \end{split}
\end{equation}
where $TP $\,--\,true positives; $FN$\,--\,false negatives; and $FP$\,--\,false positives. 


In the first set of experiments, we show the variation of adversary's success rate,
in which the $x$-axis is the
index of the devices,
and the $y$-axis is the index of precision/recall metrics.
We have several interesting observations.
First,
when the raw data has not been cleansed to remove sensitive information, the adversary can decipher the data with a high precision (see the top part of Fig. 5a).
This indicate that the adversary is often accurate in prediction.
Second, 
we can observe that all \pmb{\texttt{RASE}},
\texttt{BR}, Laplace-fisher and \texttt{BR}-mallows can effectively reduce precision to below $12.5\%$;
the minimum value
achieved is $\approx 0\%$ (see the top part of Figs. 5b, 5c, 5d and 5e).
Third,
we compare the results of \pmb{\texttt{RASE}},
\texttt{BR}, Laplace-fisher and \texttt{BR}-mallows.
Among them, \texttt{BR} exhibits a high level of precision,
which implies weaker defensive capabilities.
This is attributed to its lack of consideration for the identity confusion.
Additionally, our \pmb{\texttt{RASE}} can achieve relatively low precision across almost all sensor devices,
confirming its superiority in fortifying data robustness.

The remaining part of Fig.~\ref{fig:privacy_precision} depicts 
the performance of
\texttt{BR}, Laplace-fisher and \texttt{BR}-mallows in terms of the recall metric.
Referring back to Fig. 5a, we can observe that the true positive rate of the adversary's inference is significantly high (with average $78.5$ percent) when no sanitization is applied.
This aligns with the tendency observed in precision.
After implementing $4$ different algorithms, 
the recall metric  experiences a dramatic drop.  
For example,
\texttt{BR} achieves a success rate of $17.7$ percent on average (see the bottom part of Fig. 4b);
The performance of \texttt{BR}-mallows and 
Laplace-fisher is almost at the same level, with an averaged recall of $0.8$ percent (see the bottom part of Figs. 5c and 5d);
\pmb{\texttt{RASE}} achieves a better-case and average positive rate of $0.7 \%$ (see the bottom part of Fig. 5e). 

In the second set
of experiments,
we evaluate the performance of algorithms under different parameters.
Fig. 6a shows that the precision-recall metric decreases when device numbers increase. This suggest that the attacker's ability to accurately identify the identity labels is deteriorating.
In Fig. 6b, we presents a comparison of precision-recall under perturbation parameter $\epsilon_s$.
From the figure,
we can observe that both the precision metric and the recall metric increase as the $\epsilon_s$ increases for all schemes.
The precision-recall metric of Laplace-fisher, \texttt{BR}-mallows and our \pmb{\texttt{RASE}} are extremely close, as indicated by the coincidence of the red, orange, and purple lines.
Figs. 6c and 6d present the results  conducted on grouped disturbance using four algorithms. We note that 
the basic scheme \texttt{BR} performs poorly compared to the others.
Overall, the precision-recall curve is  overlapping for \pmb{\texttt{RASE}}, Laplace-fisher and \texttt{BR}-mallows. This is because the experiments have a limited parameter range, contributing to similar outcomes.

\noindent \textbf{Remarks.} In summary, 
both the \texttt{BR} and other shufflers can mitigate re-identification attacks. 
However,
an implication drawn from experimental results is
that shuffling can additionally
make it more challenging for adversaries to identify sensitive details.
Across all the cleansing algorithms, our \pmb{\texttt{RASE}}
stands out as it shows excellent back-tested results. 
This fact indicates that
our design of the randomize-then-shuffle paradigm
is efficient than \texttt{BR}, Laplace-fisher and \texttt{BR}-mallows.
 
\begin{figure*}[htbp]
\centering
\subfloat[]{\includegraphics[width=.24\linewidth]{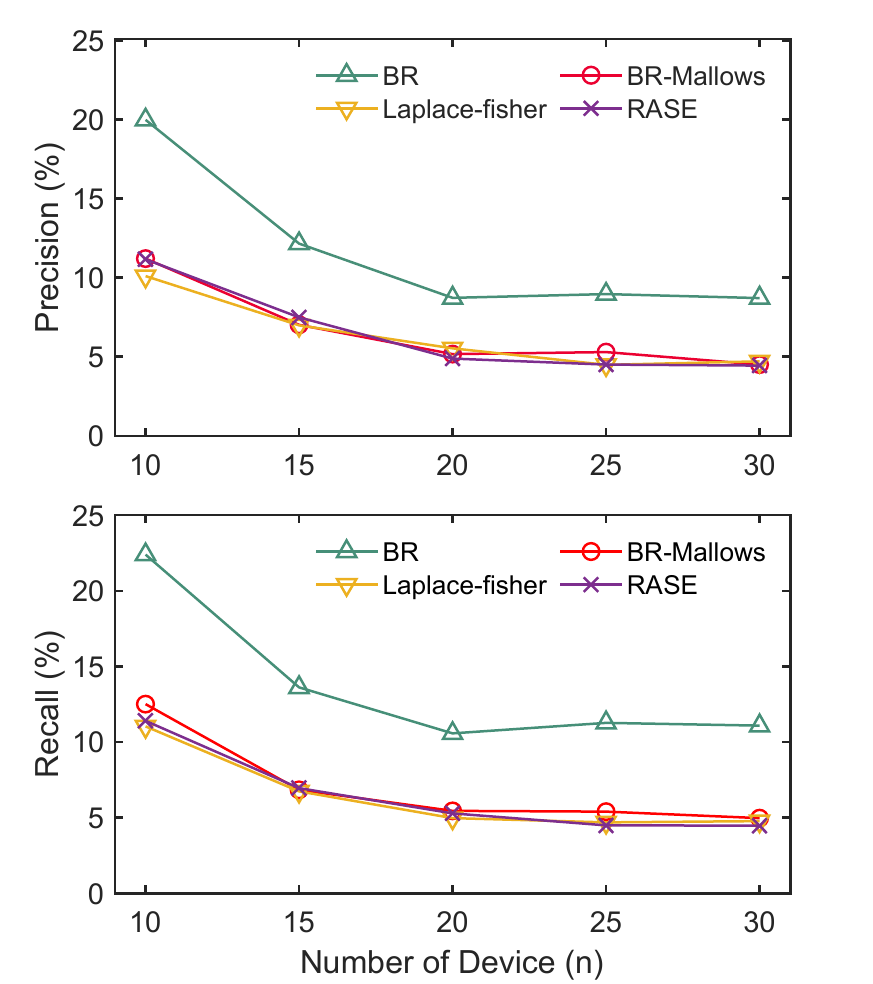}}\hspace{3pt}
\subfloat[]{\includegraphics[width=.24\linewidth]{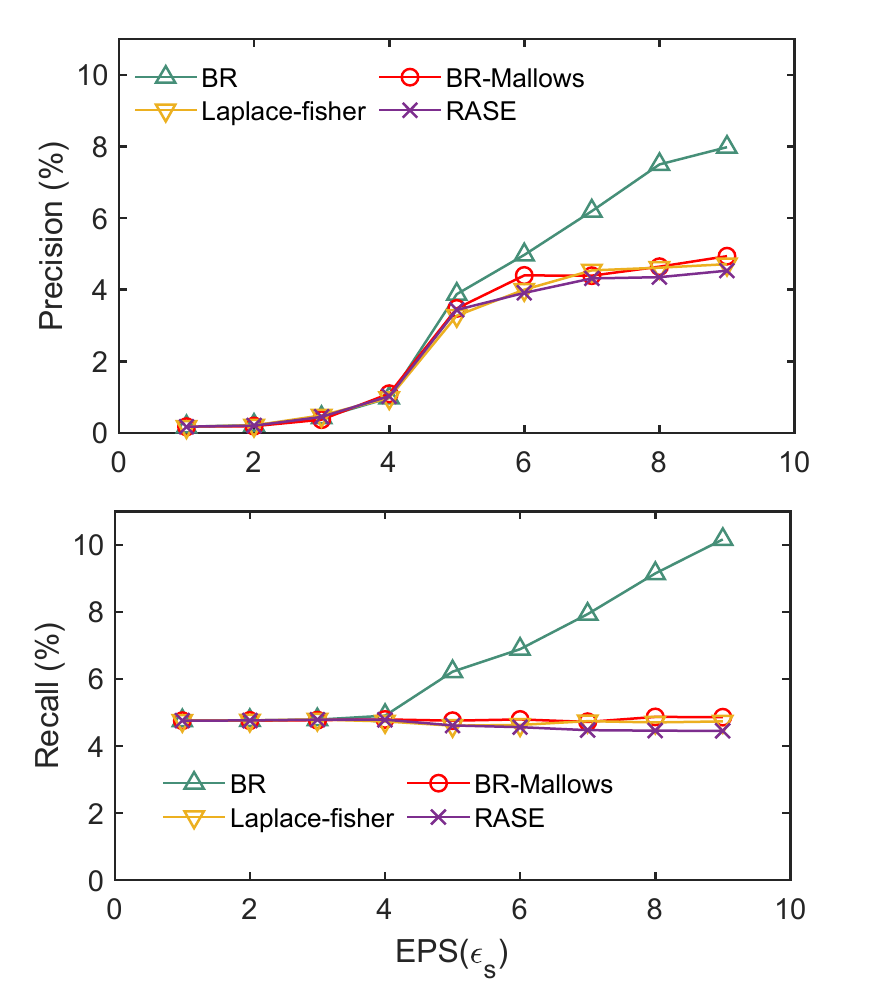}}\hspace{3pt}
\subfloat[]{\includegraphics[width=.24\linewidth]{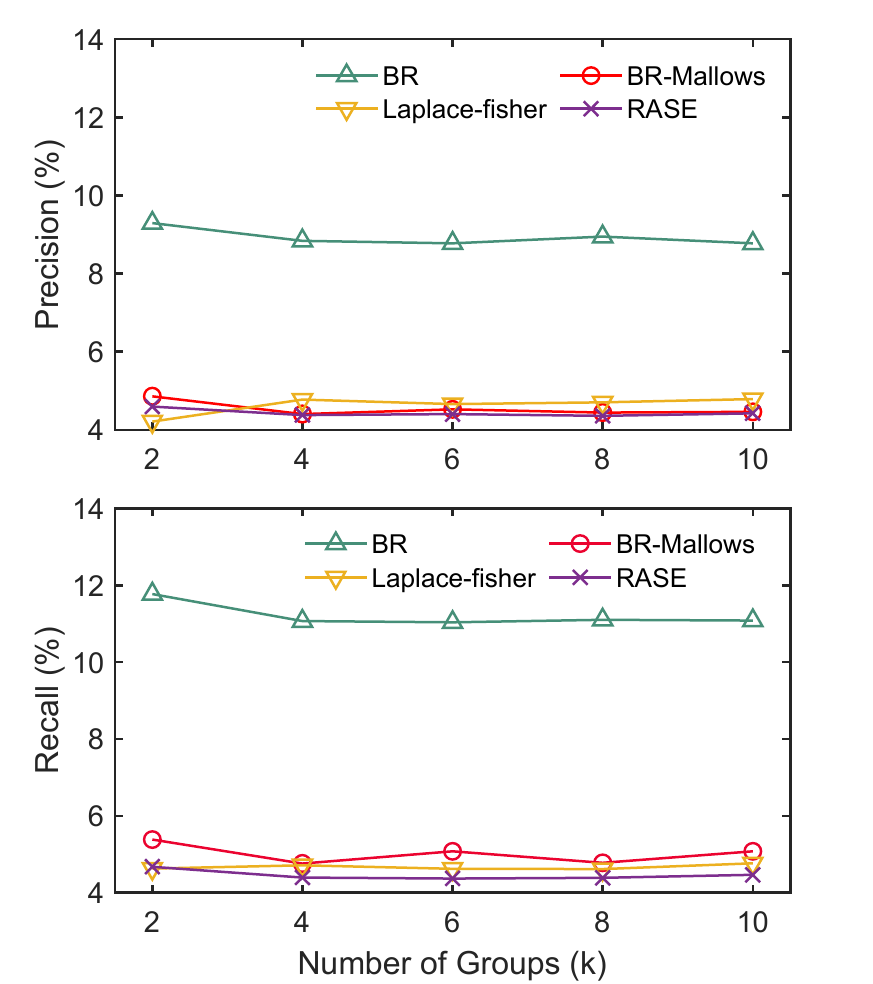}}\hspace{3pt}
\subfloat[]{\includegraphics[width=.243\linewidth]{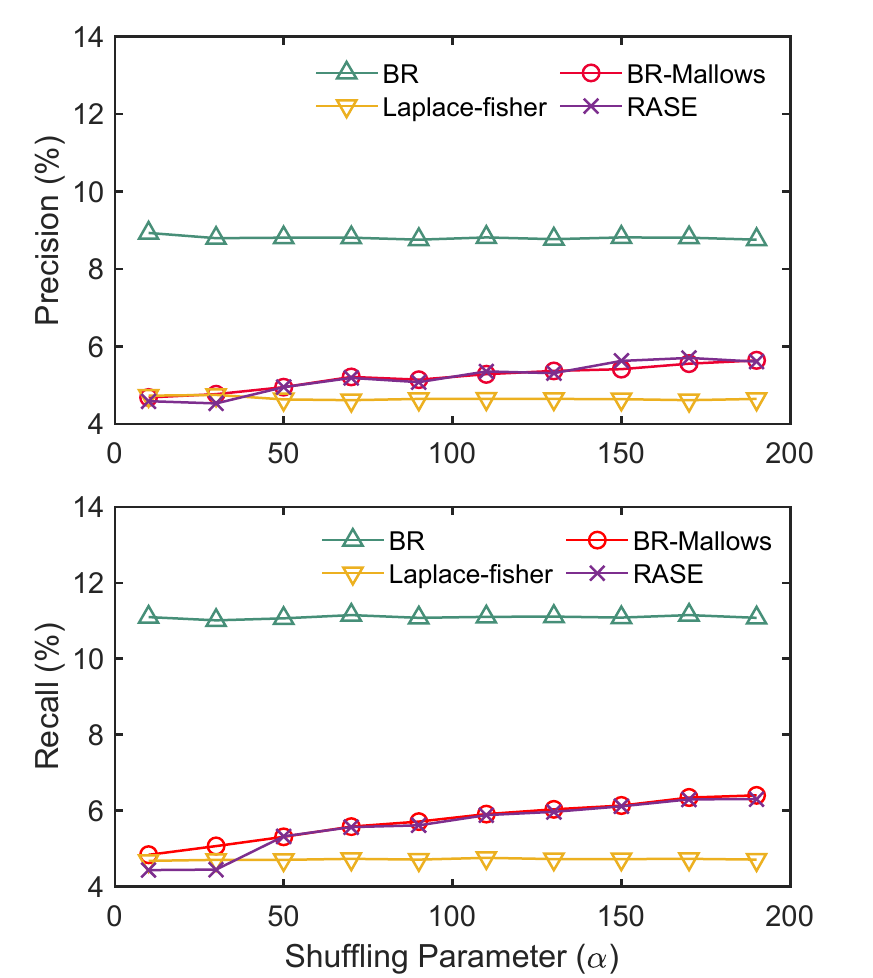}}\hspace{3pt}\caption{Analyzing the impact of factors on attack accuracy: (a) results with different numbers of devices; (b) results with varying perturbation parameters; (c) results with different numbers of groups; (d) results with varying shuffling parameters}\label{fig:precision}
\end{figure*}

\subsection{Accuracy of  Estimated Mean Values}
In this paper, we adopt two utility metrics, namely Average Aggregation Error (AAE) and Maximum
Square Error (MSE), to empirically evaluate the utility guarantee of our proposed framework.
Recall that the local randomizer imports the well-crafted stochastic noise into the raw data.
Consequently,
AAE quantifies the mean value of the error between the original data and the obfuscated data. Specifically, AAE is defined as follows:
\begin{equation}
    \begin{split}
         {\rm AAE}=\hat{\Theta}- \Theta=\frac{1}{n}{\sum}_{i=1}^n(x_i-z_i).
    \end{split}
\end{equation} 
 
On the other hand, MSE measures the maximum square difference between the estimated and true values, which provides a worst-case view of the randomizer's impact on the data. MSE is defined as:

\begin{equation}
    \begin{split}
        {\rm MSE}={\max}_{i\in \llbracket 1,n \rrbracket} (x_i-z_i)^2.
    \end{split}
\end{equation}

In the third set of experiments,
we evaluate our \pmb{\texttt{RASE}}  under different estimation methods to observe both AAE and MSE.
In addition to the average values,
we calculate the AAE and MSE for the maximum and minimum values.
Referring to Fig.~7, 
it is evident that
a larger $\epsilon_{s}$ yields 
a smaller AAE.  
Spceifically,
when $n=50$, 
the average AAE decreases from $1216.4$ to $60.9$ 
if $\epsilon_{s}$ increases from $0.1$ to $9$.
This outcome is a natural result because
the noise follows the Laplace distribution, and its magnitude is directly related to the budget $\epsilon_{s}$.
Fig.~8 shows that MSE declines when budget $\epsilon_{s}$ increases.
For example,
when $\epsilon_{s}=0.1$,
the average value of MSE is $1.7\times 10^6$,
but this value will decrease to $0.3\times 10^6$ when $\epsilon_{s}$ increases to $9$. 

Fig. 9 shows the true and perturbed load mean of smart devices over a 25-day period using our \pmb{\texttt{RASE}}. 
As expected, we can observe that while the patterns of the raw data are hidden in the perturbed data, the profile of the obfuscated data is completely different from that of the original data.
For instance, 
during one month (October 2014), 
the estimated values obtained for these estimator
all are greater than the original.
This is because that data perturbation impacts the utility of the data to some extent.
In general, 
the performance of the maximum-likelihood estimator is much better than the
other two methods. It can produce estimates that converge to the true value. 
For instance,
the actual load of the smart devices is around $187$, while the output from the maximum-likelihood estimator is approximately $213$.

\begin{figure}
  \centering
   \includegraphics[width=.45\textwidth]{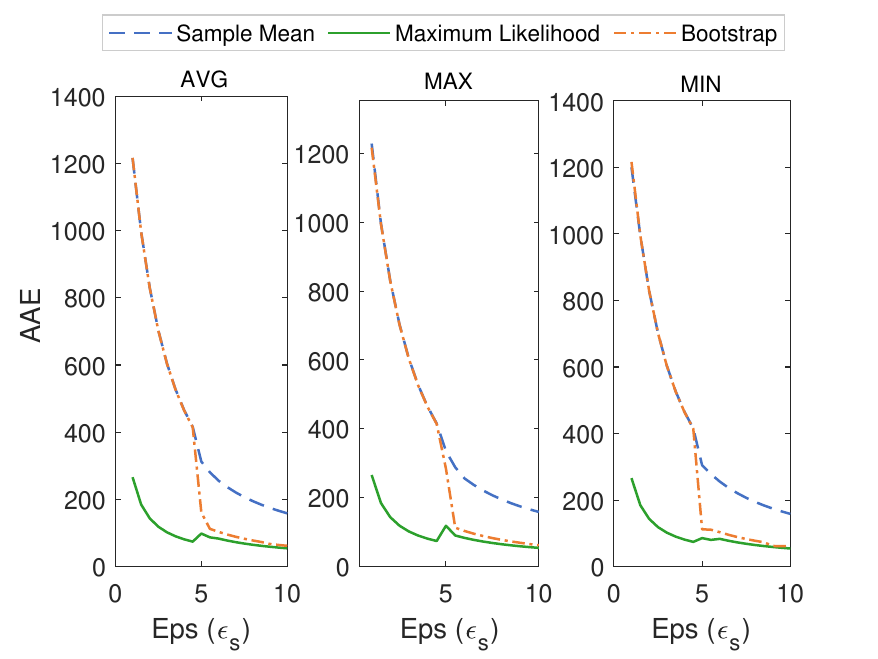}
  \caption{AAE error of the mean estimated methods.}\label{fig:l1_error}
\end{figure}

\begin{figure}
  \centering
   \includegraphics[width=.45\textwidth]{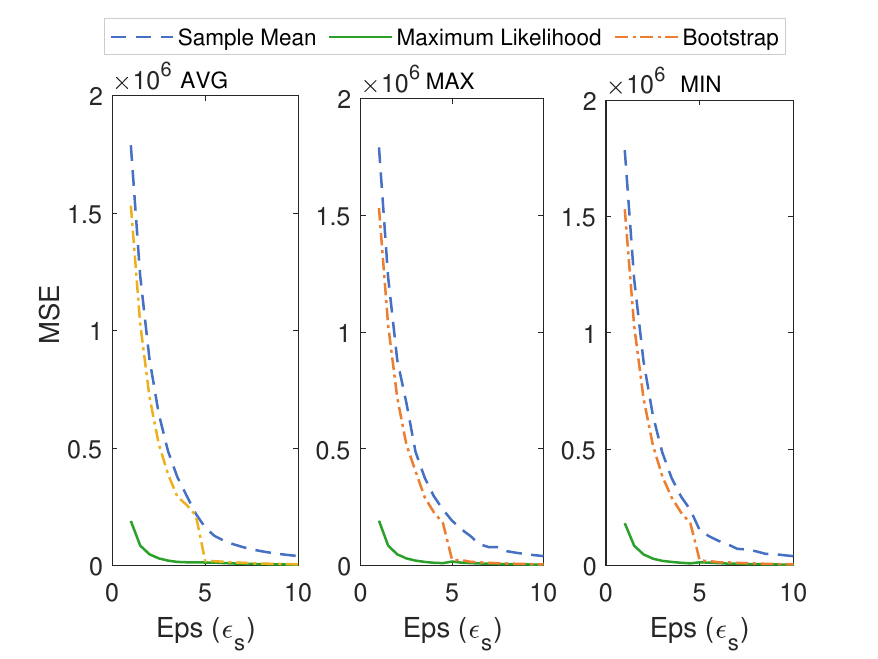}
  \caption{MAE error of the mean estimated methods.}\label{fig:l2_error}
\end{figure}


\begin{figure*}[htbp]
\centering
\subfloat[]{\includegraphics[width=.45\linewidth]{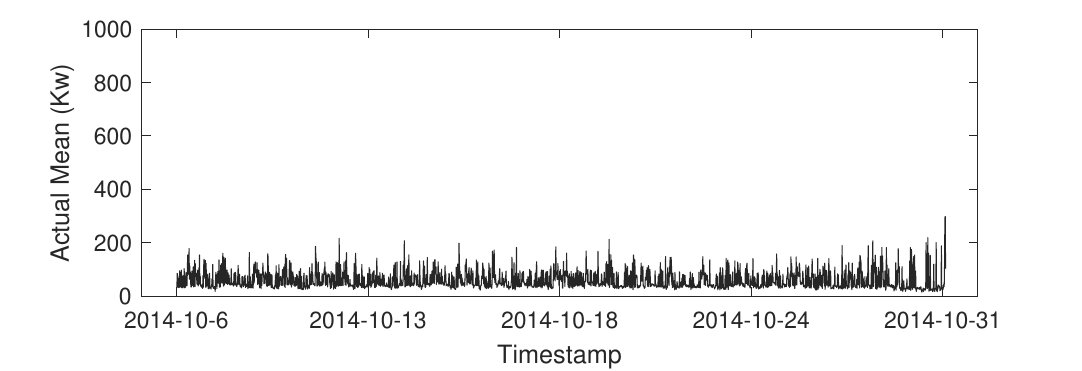}}\hspace{3pt}
\subfloat[]{\includegraphics[width=.45\linewidth]{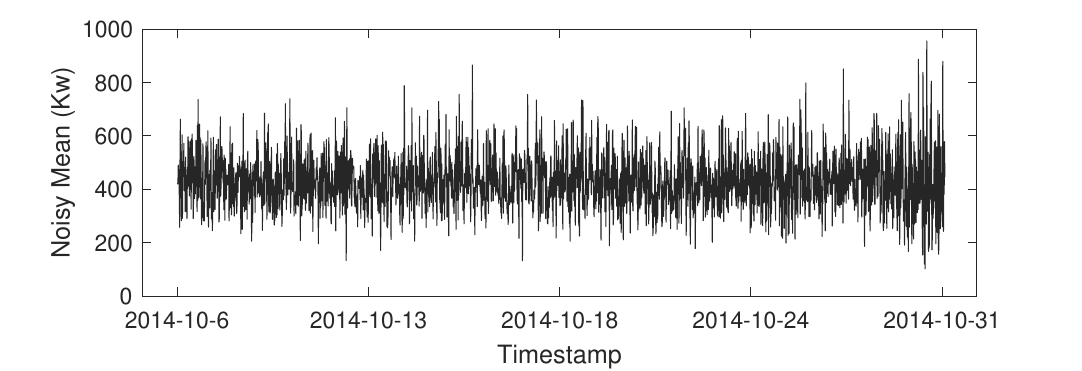}}\hspace{3pt}\\
\subfloat[]{\includegraphics[width=.45\linewidth]{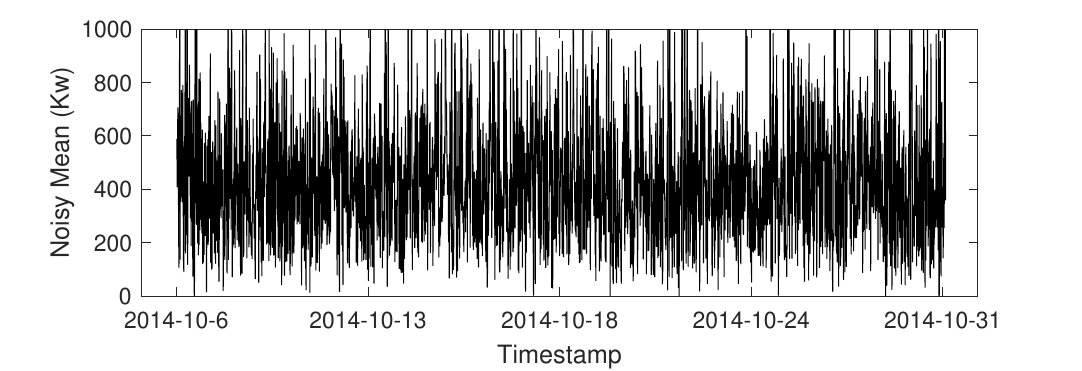}}\hspace{3pt}
\subfloat[]{\includegraphics[width=.45\linewidth]{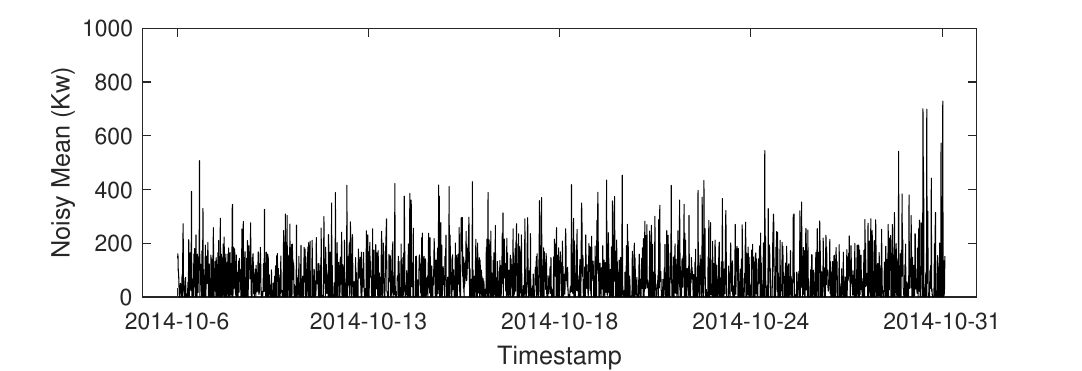}}
\caption{Observing the truth and the estimated values on 25-day traces: (a) the actual mean, (b) estimated values by sample estimator, (c) estimated values by maximum-likelihood estimator, and (d) estimated values by bootstrap estimator.} 
\end{figure*} 
\section{Related Work}
\label{sec:Related Work}
We briefly discuss the literature related to
the protection of privacy disclosure in
aggregation-based IoT applications.

\noindent \textbf{Protecting Privacy Using Anonymous Scheme.}
The conventional technology for concealing identifiable information is the anonymous scheme.
Wajahat {\it et al.}~\cite{ali2020alpha}
developed an ALPHA protocol
to enhance the untraceability of power usage data for industrial cyber-physical systems.
Its core idea is to apply orthogonal 
256-bit encryption as the secret token.
However, the practical deployment of ALPHA requires extensive resources and time.
Observing the advantages of fog computing
(e.g., latency reductionand and scalability),
Liu {\it et al.} studied a fog computing-enabled aggregation model for smart grids~\cite{liu2019enabling}.
Their model combines the double trapdoor cryptosystem with digital signatures and supports user-side function queries.
Securing the aggregation of physiological data from sensors attached to the human body is a sticky subject.
Vijayakumar {\it et al.}~\cite{vijayakumar2019efficient} proposed an anonymous authentication framework that permits the logged-in doctors to authenticate a patient’s proof-of-identity and access the status information
from the sensor devices. 
The merit of their framework is in not requiring key exchange: it assumes a trusted authority in the key distribution.
Although these solutions share our goal of hiding true identity,
their focus is different,
and our insights are much more extensive.
Specifically,  
we provide a holistic analysis for verifying the privacy,
utility and robustness
of our approach.

\noindent \textbf{Protecting Privacy Using Differential Privacy.}
In terms of performance, 
there have been some recent studies on data aggregation under differential privacy.
The notion of inference privacy
was first proposed in \cite{sun2019relationship},
where Ding {\it et al.} examined the definition of privacy coverage to 
analyze individual-level inference privacy.
Ou {\it et al.}
designed a novel data releasing mechanism, called CCDP, that 
uses the Fourier coefficients to control noise and clear up cross-correlation privacy in IoT data~\cite{ou2020optimal}.
They also shown that the CCDP mechanism satisfies $\epsilon$-Pufferfish privacy.
Cao {\it et al.} quantified the risk of differential privacy
under temporal correlations by formalizing
the privacy loss function~\cite{cao2018quantifying}.
The designed algorithms strengthen data security for time-series data leakage.
The common problem of the above research is that the utility-privacy trade-offs are not covered,
which could result in them being overprotective in some situations.

\section{Conclusion}
\label{sec:Conclusion}
In this paper,
we examine the privacy issues of IoT  information systems with respect to how
they expose to disclosure attacks when data is aggregated.
Since no published answer has touched upon the vulnerabilities before,
we presented \pmb{\texttt{RASE}}\,--\,a hybrid model to direct users to implement  privacy controls.
With the \pmb{\texttt{RASE}} model,
we developed a local randomizer that limits the output to meet the desired precision requirement, 
subsequent to adding Laplace noise.
This enables substantial utility improvement without violating the $\epsilon$-LDP guarantee.
We also developed a shuffler to provide dependable defense against identity disclosure.
To further enhance utility,
we compared three different estimators on the aggregation value of the multiple noisy data.
What is more important,
we theoretically shown that our \pmb{\texttt{RASE}} can
optimize the balance between utility loss and privacy risks.
The empirical results 
confirm our theoretical analyses 
and demonstrate the efficiency of \pmb{\texttt{RASE}}.
We believe that our work is a critical step towards
establishing stronger privacy standards for IoT data aggregation systems.
In the future,
we will investigate how to obtain more accurate estimates by exploring other promising technologies, e.g., data augmentation and machine learning.


\section*{Appendix}
\subsection{Proof of Lemma~\ref{lem:accuracy}}
\begin{proof}
We verify this by deriving the quantitative relationship between probability interval and noisy probability.
Thus, we can obtain 
\begin{equation}\label{eq:lap_CDF}
\begin{split}
&{\rm Pr}\Big[y_i\geq (1-\beta)x_i~{\rm and}~ y_i\leq(1+\beta)x_i  \Big]\\
=~& {\rm Pr}\Big[ x_i\geq y_i-\beta x_i ~{\rm and}~ x_i\leq y_i+\beta x_i\Big]\\
=~&
{\rm Pr}\Big[\eta_i\geq -\beta x_i ~{\rm and}~ \eta_i\leq \beta x_i\Big].  
\end{split}
\end{equation}

Recall that $\eta_i$ is the Laplacian random variable.
The Cumulative Distribution Function (CDF) of ${\rm Lap}(\mu,\lambda_i)$
is
\begin{equation}
F_L(o)
=
\int_{-\infty}^{o}f_L(\mu)\rm d \mu=
\begin{cases}
\frac{1}{2}{\rm exp}(-\frac{o-\mu}{\lambda_i}),& o< \mu\\
1-\frac{1}{2}{\rm exp}(-\frac{o-\mu}{\lambda_i}),& o\geq \mu
\end{cases}.\notag
\end{equation}
Utilizing the above CDF function,
we can rewrite Eq.~(\ref{eq:lap_CDF})
as
\begin{equation}\label{eq:lap_1}
\begin{split}
&{\rm Pr}\Big[\eta_i\geq -\beta x_i ~{\rm and}~ x_i\leq \beta x_i\Big]=
F_L(\beta x_i)-F_L(-\beta x_i)\\
=&1\!-\!\frac{1}{2}{\rm exp}\big(-\!\frac{\beta x_i}{\lambda_i}\big)\!-\!\frac{1}{2}{\rm exp}\big(-\!\frac{\beta x_i}{\lambda_i}\big)\geq 1\!-\!{\rm exp}\big(-\!\frac{\beta x_{max}}{\lambda_i}\big).
\end{split}\notag
\end{equation}
For a given confidence level $0 \leq \rho\leq 1$,
we have
\begin{equation}\label{eq:lap_2}
1-{\rm exp}\big(-\frac{\beta x_{max}}{\lambda_i}\big)\geq  \rho
.
\end{equation}
Also, we have
\begin{equation}\label{eq:lap_3}
    -\ln(1-p)\leq \frac{\beta x_{max}}{\lambda_i}.
\end{equation}

Substituting $\lambda_i= \frac{\Delta(x)}{\epsilon_s}$ into Eq.~(\ref{eq:lap_3}) yields the claim bound.
\end{proof}

\subsection{Proof of Theorem~\ref{theo:BR_gur}}
\begin{proof}
Let $x$, $x'$ be two adjacent scalars and $Y$ be an output.
Without loss of generality,
we assume $x>x'$ and $x- x'\leq\Delta(x)$.
Since
$\eta$ and $\eta'$
are Laplace-distributed noise  drawn from distribution ${\rm Lap}(0,\frac{\Delta(x)}{\epsilon_s})$,
$y=x+\eta$ and $y'=x'+\eta'$
are also two random variables that behave ${\rm Lap}(x,\frac{\Delta(x)}{\epsilon_s})$ and ${\rm Lap}(x',\frac{\Delta(x)}{\epsilon_s})$,
respectively.
When $\epsilon_s\geq-\frac{\Delta\cdot\ln{(1-\rho)}} {\beta x_{max}}$, the proposed mechanism \texttt{BR} returns $y$ and $y'$. 
It is easy to see that their difference only lies in mean.
Thus,
we have 
\begin{equation}\label{ineq:loc_dir}
\frac{{\rm Pr}\big{[}\texttt{BR}(x) = Y\big{]}}{ {\rm Pr}\big{[}\texttt{BR}(x')= Y\big{]}} = \frac{{\rm exp}(-\frac{\epsilon_s|y-Y|}{\Delta(x)})}{{\rm exp}(-\frac{\epsilon_s|y'-Y|}{\Delta(x)})}\leq e^{\epsilon_s}.
\end{equation}
When $\epsilon_s<-\frac{\Delta(x)\ln{(1-\rho)}} {\beta x_{max}}$, we consider all three possible case.

\noindent\textbf{Case \uppercase\expandafter{\romannumeral1}:}  if $Y=x_{min}$, 
we have 
\begin{equation}
    {\rm Pr}\big{[}\texttt{BR}(x)= x_{min}\big{]}= {\rm Pr}\big{[}y<x_{min}\big{]}.
\end{equation}

Due to $y\sim{\rm Lap}(x,\frac{\Delta(x)}{\epsilon_s})$, 
we can obtain the following:

\begin{equation}
    \begin{split}
         {\rm Pr}\big{[}y<x_{min}\big{]}
         =&\int_{-\infty}^{x_{min}} \frac{\epsilon_s}{2\Delta(x)}{\rm exp}\big({-\frac{\epsilon_s|y-x|} {\Delta(x)}}\big){\rm d} y\\
         =&\int_{-\infty}^{x_{min}} \frac{\epsilon_s}{2\Delta(x)}{\rm exp}\big({\frac{\epsilon_s(y-x)} {\Delta(x)}}\big){\rm d} y\\
         =&\frac{1}{2}\int_{-\infty}^{x_{min}} {\rm exp}\big({\frac{\epsilon_s(y-x)} {\Delta(x)}}\big) {\rm d} \big({\frac{\epsilon_s(y-x)} {\Delta(x)}}\big)\\
         =& \frac{1}{2} {\rm exp}\big({\frac{\epsilon_s(x_{min}-x)} {\Delta(x)}}\big), \notag
    \end{split}
\end{equation}
where the second equality holds because $y\leq x_{min}$,
and the third equality uses the change-of-variable theorem. 

Similarly, we have 
\begin{equation}
    {\rm Pr}\big{[}\texttt{BR}(x')= x_{min}\big{]}
         =\frac{1}{2} {\rm exp}\big({\frac{\epsilon_s(x_{min}-x')} {\Delta(x)}}\big).
\end{equation}

Thus, the ratio $\frac{{\rm Pr}\big{[}\texttt{BR}(x)= x_{min}\big{]}}{{\rm Pr}\big{[}\texttt{BR}(x')= x_{min}\big{]}}=e^{{\frac{\epsilon_s(x'-x)} {\Delta(x)}}}\leq e^{-\epsilon_s}\leq e^{\epsilon_s}$.

\noindent\textbf{Case \uppercase\expandafter{\romannumeral2}:} if $Y=x_{max}$, 
we have $\frac{{\rm Pr}\big{[}\texttt{BR}(x)= x_{max}\big{]}}{{\rm Pr}\big{[}\texttt{BR}(x')= x_{max}\big{]}}=\frac{{\rm Pr}\big{[}y>x_{max}\big{]}}{{\rm Pr}\big{[}y'>x_{max}\big{]}}\leq e^{-\epsilon_s}\leq e^{\epsilon_s}$.
The correctness of this equation can be verified analogously as that in Case \uppercase\expandafter{\romannumeral1}.

\noindent\textbf{Case \uppercase\expandafter{\romannumeral3}:} if $Y$ falls in the range
$[x_{min}, x_{max}]$,
the
output is generated by injecting only the Laplace noise,
without the clamping step. 
By Eq.~(\ref{ineq:loc_dir}), we can obtain $\frac{{\rm Pr}\big{[}\texttt{BR}(x) = Y\big{]}}{ {\rm Pr}\big{[}\texttt{BR}(x')= Y\big{]}} \leq e^{\epsilon_s}$. 

Combining these cases,
we conclude that \texttt{BR} is $\epsilon_s$-LDP.
\end{proof}

\subsection{Proof of Theorem~\ref{theo:Np}}
\begin{proof} 
We prove this by presenting a reduction from the NP-hard balanced clustering problem~\cite{aloise2018sampling}.
Let $\Omega$ be a set of points in the plane
and let $f_c$ be a function defined on pairs of points.
The balanced clustering problem is 
to partition $\Omega$ into $P$ groups $C_1$, $C_2$, $\ldots$, $C_P$,
such that the intra-group unsimilarity $\max\{ \max \{f_c(l,t): l,t\in C_p\} : 1\leq p \leq P\}$
is minimized.
Let us go back to the GRP problem.
Since the sensitivity of Kendall-tau distance is an increasing function of the width (see Eq.~(\ref{eq:kd_sens})),
minimizing $\omega_{\xi}$ minimizes $\Delta(\sigma_0:\, d_K,\, \xi)$.
We can rewrite the target function as $\min \max_{G_l\in \xi} \omega_{\xi}$ among the $k$ groups.
It is easy to find that the modified problem is identical to the balanced clustering problem.
Therefore, the GRP problem is NP-hard too.
\end{proof}

\subsection{Proof of Lemma~\ref{lem:shuff}}
\begin{proof}
Given a set of masked inputs $[y]=\{y_1, \ldots, y_n\}$,
the view of the shuffler in our framework is a permutation:
$\sigma_{\vec{y}}=(v_1,  v_2, \ldots, v_n)$,
where $v_1={\rm Idx}(y_1)$, $v_2={\rm Idx}(y_2)$ and so on.
At some iteration $i$,
the mechanism chooses a number $j\in\llbracket i+1,n \rrbracket$ at random,
and swaps $v_i$ and $v_j$.
To do so,
the permutation changes to
\begin{equation}
\sigma'_{\vec{y}}=\big(v_1, \ldots, v_{i-1}, v_j, v_{i+1},\ldots v_{j-1}, v_i, v_{j+1}, \ldots, v_n\big).\notag
\end{equation}

After interchanging with $v_i$,
the item $v_j$ remains unmoved for all subsequent iterations.
By the definition of function ${\rm Idx}(\cdot)$,
we have $n-i\geq |v_i-v_j|\geq 1$ for $\forall i\in \llbracket 1,n-1 \rrbracket$.

Assume that the adversary knows 
not only the output permutation
but also the way to 
reconstruct the final permutation.
Concretely,
the probabilities of choosing
a item correctly are $\frac{1}{n-1}, \frac{1}{n-2}, \frac{1}{n-3}, \ldots, 1$, respectively.
The lemma then follows.
\end{proof}

\subsection{Proof of Theorem~\ref{theo:d_privacy}}
\begin{proof}
    To confirm this, we will need to show that 
    the auxiliary shuffler also satisfies the $(\alpha, \xi)$-$d_{\sigma}$ privacy. 
    Let $\texttt{RS}_{aux}$ be the auxiliary shuffling mechanism. 
    If we can get that $\texttt{RS}_{aux}$ 
    produces $(n-1)!$ different permutations,
    then by Lemma~\ref{lem:shuff},
    we can know that it assigns every permutation an equal probability.
    That is, for any pair of permutations $\sigma$ and $\sigma'$, we have $\forall \tau$, 
    \begin{equation}
        {{\rm Pr}\big{[}\texttt{RS}_{aux}(\sigma)= \tau\big{]}}={{\rm Pr}\big{[}\texttt{RS}_{aux}(\sigma')= \tau\big{]}}.
    \end{equation}

    
    To get that,
    we consider a permutation of $n$ elements.
    In the initial step, 
    our $\texttt{RS}_{aux}$ can pick the random integer in the range $\llbracket2, n\rrbracket$,
    which means that
    there are $n-1$ possible options.
    $\texttt{RS}_{aux}$ will do $n-1$ iterations, 
    reducing the range from $n-1$ to $n - (n-1) = 1$.
    Therefore, 
    the number of permutations that 
    $\texttt{RS}_{aux}$ generates is $(n-1)!$.
    This establishes the proof.   
\end{proof}

\subsection{Proof of Theorem~\ref{theo:MAE}}
\begin{proof}
In order to show that the theorem holds in fact, 
we will need a technical lemma from~\cite{Dwork2013The}.
\begin{lem}[Dwork {\it et al.}~\cite{Dwork2013The}]\label{lem:Lap_MAE}
    The expected maximum error of the Laplace mechanism is $O(\frac{1}{\epsilon_s})$.
\end{lem}

Consider calculating the average of a anonymized set $\left[z\right]=\left \{ z_1, \ldots, z_n\right \}$,
whose
elements are either all unclamped or all clamped,
depending on the accuracy requirement. 
In the former situation,
each data $z_i$,
$i\in \llbracket n \rrbracket$, is
perturbed using ${\rm Lap}\left(\Delta(x)/\epsilon_s\right)$; thus the mean error corresponds to Lemma~\ref{lem:Lap_MAE}.
Let $\left[x\right]=\left \{x_1, \ldots, x_n\right \}$ be the original data set.
Then 
the magnitude change in the data when clamping occurs at a randomizer is bounded by $\left[x_{min},x_{max}\right]$.
That is,
any output only increases the error by
up to $\left|x_{max}-x_{min} \right|$.

Thus,
we have 
\begin{equation}
    \begin{split}
   \mathbbm{E}\big\|\hat{\Theta}_{SE}- \Theta\big\|_1= &\mathbbm{E}\Bigg[\left|\frac{1}{n}{\sum}_{i\in\llbracket 1,n \rrbracket} z_i-\frac{1}{n}{\sum}_{i\in\llbracket 1,n \rrbracket}  x_i \right|\Bigg]\\
   \leq &\frac{1}{n}{\sum}_{i\in\llbracket 1,n \rrbracket}  \mathbbm{E}\big[\left|z_i-x_i\right|\big]\\
       \in & \left \{O\left(\frac{1}{\epsilon_s}\right), \Delta(x)\right \} ~. 
    \end{split}
\end{equation}  
    This establishes the proof.
\end{proof}

\bibliographystyle{IEEEtran}
\bibliography{reference}

\end{document}